\documentclass[letterpaper,useAMS,usenatbib]{mn2e}
\usepackage{graphicx}
\usepackage{xcolor}
\usepackage{amssymb}
\usepackage{amsmath}	

\def\ms{\mbox{$M_{\star}$}}
\def\mh{\mbox{$M_{\rm 200c}$}}   
\def\mstar {h^{-1}~\rm M_{\sun}}
\def\mhalo {h^{-1}~\rm M_{\sun}}
\def\msun {\rm M_{\sun}}
\def\mpc {h^{-1}~\rm Mpc}
\def\sag {\textsc {sag}}
\def\mdsag {\textsc {md$\_$sag}}
\def\mdsagbeta {\textsc {sag$_{\beta1.3}$}}
\def\tng {{\sc IllustrisTNG300}}

\title[Galactic conformity at few Mpc scales]{On the environmental influence of groups and clusters of galaxies beyond the virial radius: Galactic conformity at few Mpc scales}
\author[Lacerna et al.]
  {Ivan Lacerna,$^{1, 2}$\thanks{E-mail: ivan.lacerna@uda.cl} Facundo Rodriguez,$^{3, 4}$ Antonio~D. Montero-Dorta,$^{5}$ \newauthor 
  Ana L. O'Mill,$^{3, 4}$
  Sofía A. Cora,$^{6,7}$
  M.~Celeste Artale,$^{8,9}$  
  Andrés N. Ruiz,$^{3, 4}$ 
    \newauthor Tomás Hough$^{6}$ and Cristian~A. Vega-Mart\'inez$^{10,11}$ \\
    $^1$Instituto de Astronom\'ia y Ciencias Planetarias, Universidad de Atacama, Copayapu 485, Copiap\'o, Chile\\
    $^2$Millennium Institute of Astrophysics, Nuncio Monsenor Sotero Sanz 100, Of. 104, Providencia, Santiago, Chile \\
    $^3$Instituto de Astronomía Teórica y Experimental, CONICET-UNC, Laprida 854, X5000BGR, Córdoba, Argentina\\
    $^4$Observatorio Astronómico de Córdoba, UNC, Laprida 854, X5000BGR, Córdoba, Argentina.\\
	$^5$Departamento de F\'isica, Universidad T\'ecnica Federico Santa Mar\'ia, Casilla 110-V, Avda. Espa\~na 1680, Valpara\'iso, Chile\\    
$^{6}$Instituto de Astrof\'isica de La Plata (CCT La Plata, CONICET, UNLP), Observatorio Astron\'omico, Paseo del Bosque
B1900FWA, \\ La Plata, Argentina\\
$^{7}$Facultad de Ciencias Astron\'omicas y Geof\'isicas, Universidad Nacional de La Plata, Observatorio Astron\'omico, Paseo del Bosque, \\
B1900FWA La Plata, Argentina \\	
	$^8$Institut f\"ur Astro- und Teilchenphysik, Universit\"at Innsbruck, Technikerstra\ss e 25/8, A-6020 Innsbruck, Austria\\
	$^{9}${Department of Physics and Astronomy, Purdue University, 525 Northwestern Avenue, West Lafayette, IN 47907, USA}\\
	$^{10}$ Instituto de Investigaci\'on Multidisciplinar en Ciencia y Tecnolog\'ia, Universidad de La Serena, Ra\'ul Bitr\'an 1305, La Serena, Chile\\
    $^{11}$ Departamento de Astronom\'ia, Universidad de La Serena, Av. Juan Cisternas 1200 Norte, La Serena, Chile\\
	}   

\pagerange{\pageref{firstpage}--\pageref{lastpage}} \pubyear{000}

\begin{document}
\label{firstpage}

\maketitle
\begin{abstract} 
The environment within dark matter haloes can quench the star formation of galaxies. However, environmental effects beyond the virial radius of haloes ($\gtrsim$ 1 Mpc) are less evident. An example is the debated correlation between colour or star formation in central galaxies and neighbour galaxies in adjacent haloes at large separations of several Mpc, referred to as two-halo galactic conformity.  
We use two galaxy catalogues generated from different versions of 
the semi-analytic model {\sc sag} applied to the {\sc mdpl2} cosmological simulation 
and the \tng\ cosmological hydrodynamical simulation to study the two-halo conformity by measuring the quenched fraction of neighbouring galaxies as a function of the real-space distance from central galaxies. 
We find that low-mass central galaxies in the vicinity of massive systems 
(\mh\ $\geq$ 10$^{13}$ $\mhalo$) 
out to 5 $\mpc$ are preferentially quenched compared to other central galaxies at fixed stellar mass \ms\ or fixed host halo mass \mh\ at $z \sim 0$. In all the galaxy catalogues is consistent 
that the low-mass (\ms\ $<$ 10$^{10}$ $\mstar$ or \mh\ $<$ 10$^{11.8}$ $\mhalo$) central galaxies in the vicinity of clusters and, especially, groups of galaxies mostly produce the two-halo galactic conformity. 
On average, the quenched low-mass central galaxies are much closer to massive haloes than star-forming central galaxies of the same mass (by a factor of $\sim$ 5). 
Our results agree with other works regarding
the environmental influence of massive haloes that can extend beyond the virial radius and affect nearby low-mass central galaxies.
\end{abstract}

\section{Introduction}

The description of the dependence of the physical properties of galaxies on their environment is paramount to understand galaxy formation. 
Galaxies residing within groups and clusters are strongly affected by different processes that modify their gas content,
e.g., ram-pressure stripping
\citep[]{GG1972, GH1985, Aragon-Salamanca+1993, Solanes+2001, McCarthy+2008, Cortese+2011, Gavazii+2018, LinL+2019, Roberts+2019, Schaefer+2019}, starvation or `strangulation'
\citep{Larson+1980,  Peng+2015, Spindler+2018, Garling+2020},
and high-speed galaxy encounters or `galaxy harassment' 
\citep{Moore+1996, Moore+1998, LinL+2010}.

Interest in environmental effects at large scales
($\gtrsim$ 1 Mpc), typically well beyond the virial radius of galaxy groups and clusters, has increased in the last few years
\citep[e.g.,][]{Wetzel+2012,Bahe+2013,Benitez-Llambay+2013,Cybulski+2014,Campbell+2015,Hearin+2015,Bahe+2017,Goddard+2017_465env,ZhengZheng+2017,Zinger+2018,Duckworth+2019,Kraljic+2019,Pallero+2019,Tremmel+2019,ZhengZheng+2019,Pandey&Sarkar2020,ZhangY+2021}.
The effect of the large-scale environment 
on massive galaxies is less strong than 
the effect of mass \citep[e.g.][]{Alpaslan+2015},
but seems to play a role in less massive galaxies 
\citep{Peng+2010full, Bluck+2014, AF+2018}.
The minimum role of the large-scale environment at higher masses is likely due to the dominant presence of active galactic nuclei (AGN) feedback in these galaxies
\citep[e.g.][]{Bower+2006, Hirschmann+2013, Bluck+2016, Guo+2019}.
Environmental effects also depend on the orbital evolution of galaxies. 
Galaxies found beyond the virial radius of the cluster ($1 - 2\, {\rm Mpc}$ from the cluster centre) might be recent infallers or galaxies that have passed once near the cluster centre, where environmental effects are stronger and are in their way out of the cluster.
Galaxies that experienced this latter phenomenon are known as
“backsplash”
galaxies \citep{Gill+2005, Pimbblet2011, MurielCoenda2014, Haggar+2020}.
About 
$60$ per cent
of galaxies in the 
region between one and two virial radii around a galaxy cluster would be backsplash galaxies \citep{Haggar+2020}.

A remarkable case of the effect of the environment at different scales is galactic conformity 
\citep[e.g.][]{Weinmann+2006, Kauffmann+2013, Phillips+2014, Kauffmann2015, Knobel+2015, Paranjape+2015, Bray+2016, Hearin+2016, LeeH+2016, Berti+2017, Sin+2017, Calderon+2018, Lacerna+2018MNRAS, RafieferantsoaDave2018, Sun+2018, Tinker+2018, Treyer+2018, ZuMandelbaum2018, Sin+2019, Alam+2020, Otter+2020,LiL+2021,Maier+2022}.
This term is 
used to describe the observed correlation between colour or star formation activity in central galaxies 
and their satellite galaxies. 
Observationally, central galaxies are usually 
identified
in the centre of galaxy groups or clusters
or as galaxies without relatively bright neighbours;
theoretically, they reside near the centre of the potential well of host dark matter haloes.
\citet{Weinmann+2006} defined the term galactic conformity after finding that quenched central galaxies have a higher fraction of quenched satellite galaxies compared to star-forming central galaxies in galaxy groups of similar mass at $z$ $<$ 0.05. Later, \citet{Kauffmann+2013} found a galactic conformity effect between low-mass central galaxies with low specific star formation rate (sSFR) or gas content and neighbour galaxies with low sSFR out to scales of 4 Mpc at $z$ $<$ 0.03. %
These results motivated the distinction between the conformity measured at small separations between the central galaxy and their satellite galaxies within a dark matter halo and the signal measured at large separations of several Mpc between the central galaxy and neighbour galaxies in adjacent haloes. The former is referred to as one-halo conformity, while the latter
is called two-halo conformity.

Cosmological hydrodynamical simulations 
\citep[e.g.][]{Bray+2016}, 
semi-analytic models of galaxy formation \citep[e.g.][]{Lacerna+2018MNRAS}, and mock galaxy catalogues \citep[e.g.][]{Sin+2017, Tinker+2018} 
show
a correlation in colour or star formation between central (primary) galaxies and neighbour (secondary) galaxies at Mpc scales, i.e., two-halo conformity. However, the signal is smaller compared 
to
observations because the latter use isolation criteria to select primary galaxies that include a small fraction of satellite galaxies. 
The overall two-halo conformity decreases when only central galaxies are considered in the selection of the primaries \citep{Bray+2016, Sin+2017, Lacerna+2018MNRAS, Tinker+2018}.
Furthermore, \citet{Sin+2017} found that the two-halo conformity out to projected distances of 3--4 Mpc from central galaxies in mock catalogues is primarily related to the environmental influence of very large neighbouring haloes.
\cite{Zinger+2018} suggest that the scenario of satellite quenching in the environment of galaxy clusters, which extends to $\sim$2–3 virial radii, is consistent with the galactic conformity over large distance scales
of several Mpc. On the other hand,
\citet{Lacerna+2018MNRAS} found that the two-halo conformity is only detected for central galaxies in relatively low-mass haloes ($M_{\rm halo}$ $\le$ 10$^{12.4}$ $\mpc$). It has been shown that relatively massive haloes could disrupt the average growth of near smaller objects \citep[e.g.][]{WangH+2007, Dalal+2008, Hahn+2009, Behroozi+2014}
and, therefore, affect their properties \citep[e.g.][]{LacernaPadilla2011,Salcedo+2018,MansfieldKravtsov2020}.
Thus, the two-halo conformity may result from galaxies hosted by low-mass haloes 
affected by nearby massive systems.

This paper uses cosmological numerical simulations to 
extend previous works about
galaxies that are quenched preferentially in the infall region around dense and massive structures 
in the local Universe. 
In particular,
we study whether the two-halo conformity at a few Mpc scales is given by central galaxies in the vicinity of galaxy groups and clusters.

The outline of the paper is as follows. Section \ref{sec_data} describes the cosmological simulations and synthetic galaxy catalogues used in this paper. The methodology to measure galactic conformity is presented in Section \ref{sec_method}. The results with the correlations of sSFR 
are shown in Section \ref{sec_results}. We discuss our results in Section \ref{sec_discussion} and the conclusions are given in Section \ref{sec_conclusion}.

The cosmological simulations in this paper use different values of the reduced Hubble constant, $h$, 
defined as $H_0 = 100$ $h$ km s$^{-1}$ Mpc$^{-1}$.
We opted for scaling $h$ explicitly
throughout this paper 
with the following dependencies
unless the 
value of $h$ is specified:
 stellar mass and halo mass in $\mstar$, physical scale in $\mpc$,  
 and sSFR
 in $h$ yr$^{-1}$.

\section{Cosmological simulations}
\label{sec_data}

We use cosmological numerical simulations of big volumes that contain a large number of galaxy groups and clusters to obtain statistically significant results, and with a good mass resolution to sample galaxies with stellar masses above 
$7 \times 10^8$ $\mstar$
as well. We study the conformity at few Mpc scales using three different 
galaxy catalogues:
two generated by applying a semi-analytic model (SAM) of galaxy formation  and evolution
to a large {\em N}-body simulation of dark matter (Sect. \ref{sec_SAG}), and 
other extracted from
a cosmological hydrodynamical simulation (Sect. \ref{sec_TNG}). 

It is worth noticing that the scope of this paper is not 
to compare results from
SAMs and hydrodynamical simulations, but to study a possible excess of correlation in star formation and colour out to scales of few Mpc 
as a result of the presence of
central galaxies in the vicinity of relatively massive systems using different types of galaxy formation models.
In this regard, this paper does not have a simulated fiducial catalogue.

\subsection{SAG galaxy catalogues}
\label{sec_SAG}

We analyse two galaxy catalogues generated by applying the SAM \sag~to the dark matter only MultiDark Planck 2 ({\sc mdpl2}) cosmological simulation 
\citep{Klypin+2016, Knebe+2018}.
The model \sag~originates from the SAM described
in \citet{Springel+2001} and was subsequently modified as detailed
in \citet{Cora2006},
\citet{LCP2008}, \citet{Tecce+2010}, 
\citet{Orsi+2014}, \citet{MunnozArancibia+2015},
\citet{Gargiulo+2015} and \citet{Cora+2018}; the latter work presents the latest version of the model.
The {\sc mdpl2} has a huge volume of (1 $h^{-1}$ Gpc)$^3$ and dark matter mass resolution of $1.5 \times 10^9\,\mhalo$.
It is consistent with
a flat $\Lambda$CDM model characterized by Planck cosmological parameters:
$\Omega_{\rm m}$~=~0.307, $\Omega_\Lambda$~=~0.693, 
$\Omega_{\rm B}$~=~0.048, $n_{\rm s}$~=~0.96
and $H_0$~=~100~$h^{-1}$~km~s$^{-1}$~Mpc$^{-1}$, where $h$~=~0.678 
\citep{Planck+2013}.
Dark matter haloes have been identified with the
\textsc{Rockstar} halo finder
\citep{Behroozi_rockstar}, and merger trees were constructed with 
\textsc{ConsistentTrees} \citep{Behroozi_ctrees}.
 
The halo catalogues and merger trees
constitute the input of the model \sag~which assigns 
one galaxy to each new detected halo in the simulation to generate the galaxy population. Central galaxies reside within main host haloes detected over the background density. Those haloes lying within another dark matter halo are subhaloes and contain satellite galaxies. 
Those galaxies that are assigned to dark matter subhaloes that are no longer
identified by the halo finder (either because they have been disrupted by tidal effects or merged with the main host halo, or simply because of resolution effects of the underlying simulation) are called orphan satellites, and their orbital evolution is tracked semi-analytically in a pre-processing step before applying \sag~to the dark matter only simulation \citep{Cora+2018,Delfino+2022}.\footnote{The orbital evolution model described in \citet{Delfino+2022} considers a NFW density profile \citep{NFW1997} for both the host halo and the unresolved subhalo. Here, a previous version of the model is used assuming an isothermal density profile for both cases.}

The evolution of galaxy properties is tracked by \sag~considering a set of physical processes that regulate the circulation of mass and metals among the different baryonic components of the galaxy (hot gas halo, gaseous and stellar discs, stellar bulge), namely, 
radiative cooling of the hot halo gas, star formation 
(quiescent and in starbursts triggered by mergers and disc instabilities), 
chemical enrichment produced by 
stellar winds and different types of 
supernovae, feedback from supernovae and from active galactic nuclei, and environmental effects such as tidal stripping and ram pressure stripping. In particular, the value of ram pressure at the radial position of the satellite galaxies is obtained from an analytic profile that depends on halo mass and redshift, obtained by fitting the information provided by hydrodynamical simulations of groups and clusters of galaxies \citep{VegaMartinez+2022}. 
Ram pressure exerted over satellite galaxies removes their hot gas gradually after infall.
When the ratio between the hot gas mass and the baryonic mass 
of a satellite decreases below $0.1$, ram pressure can strip 
gas
from the galaxy disc.
The implementations of all these processes involve free parameters that have been calibrated to a set of observed 
relations of galaxy properties by using the 
\textit{Particle Swarm Optimisation} technique \citep{Ruiz+2015}.

The two galaxy catalogues used in this study have been generated with the version of \sag~previously described, and differ only in the value of the parameter $\beta$ involved in the explicit redshift dependence of the reheated and ejected mass by supernovae feedback (see equation 10 and 12 of \citealt{Cora+2018}),
which is based on relations measured from 
full-physics hydrodynamical simulations.
One of the catalogues is characterized by a value given by the calibration process ($\beta=1.99$), while the other was generated by adopting a smaller value ($\beta=1.3$) in order to achieve better consistency with the observational trends followed by the fraction of local passive  satellites as a function of stellar mass, halo mass, and the halo-centric distances \citep{Cora+2018}. However, the larger value of $\beta$ allows to reproduce the evolution of the  the mass-metallicity relation of galaxies in the redshift range $0<z<3.5$ \citep{Collacchioni+2018}.
We refer to the galaxy catalogues with the larger and smaller values of $\beta$ as \mdsag~and \mdsagbeta, respectively.

\begin{figure*}
\includegraphics[width=\columnwidth]{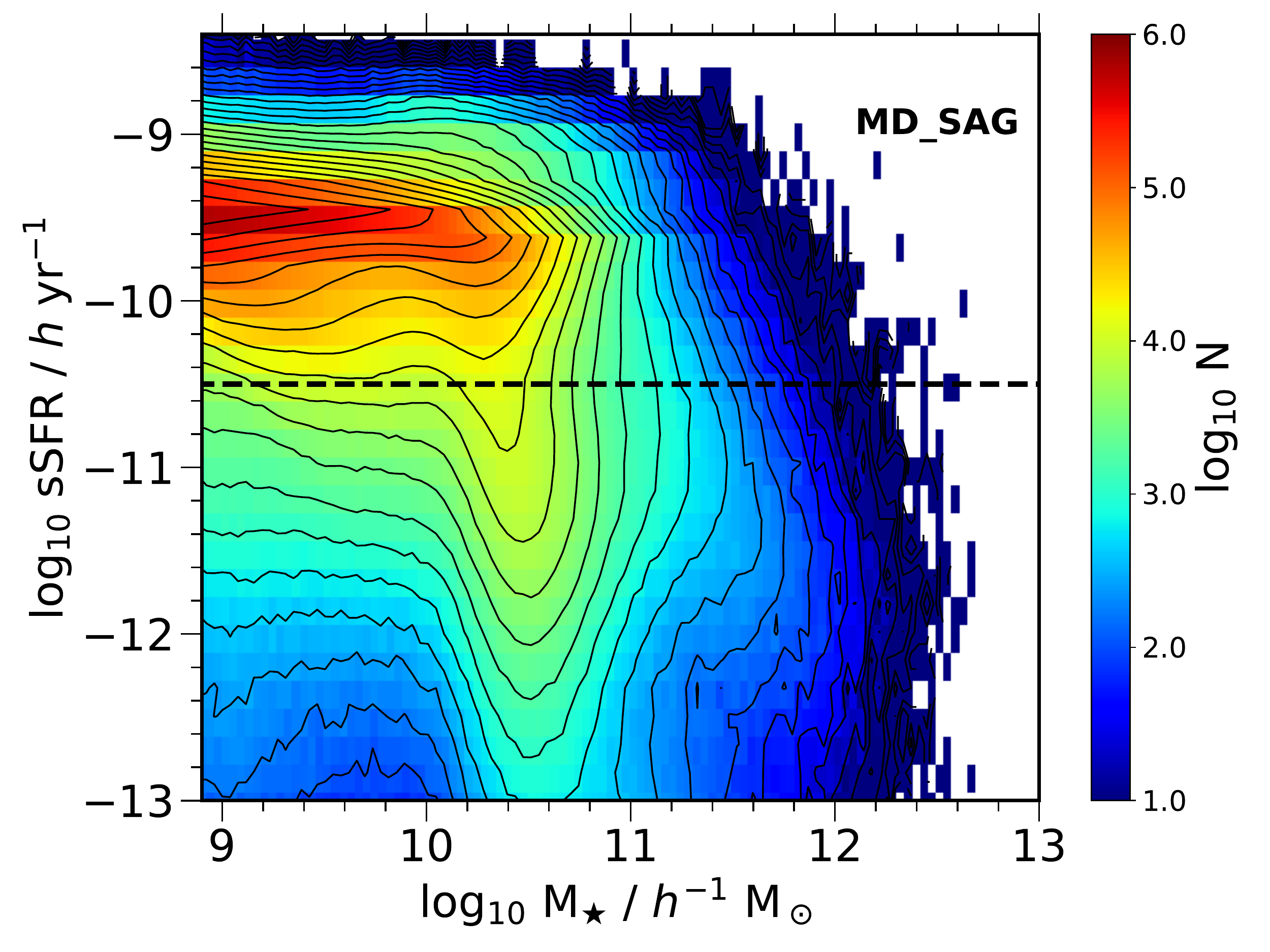}
\includegraphics[width=\columnwidth]{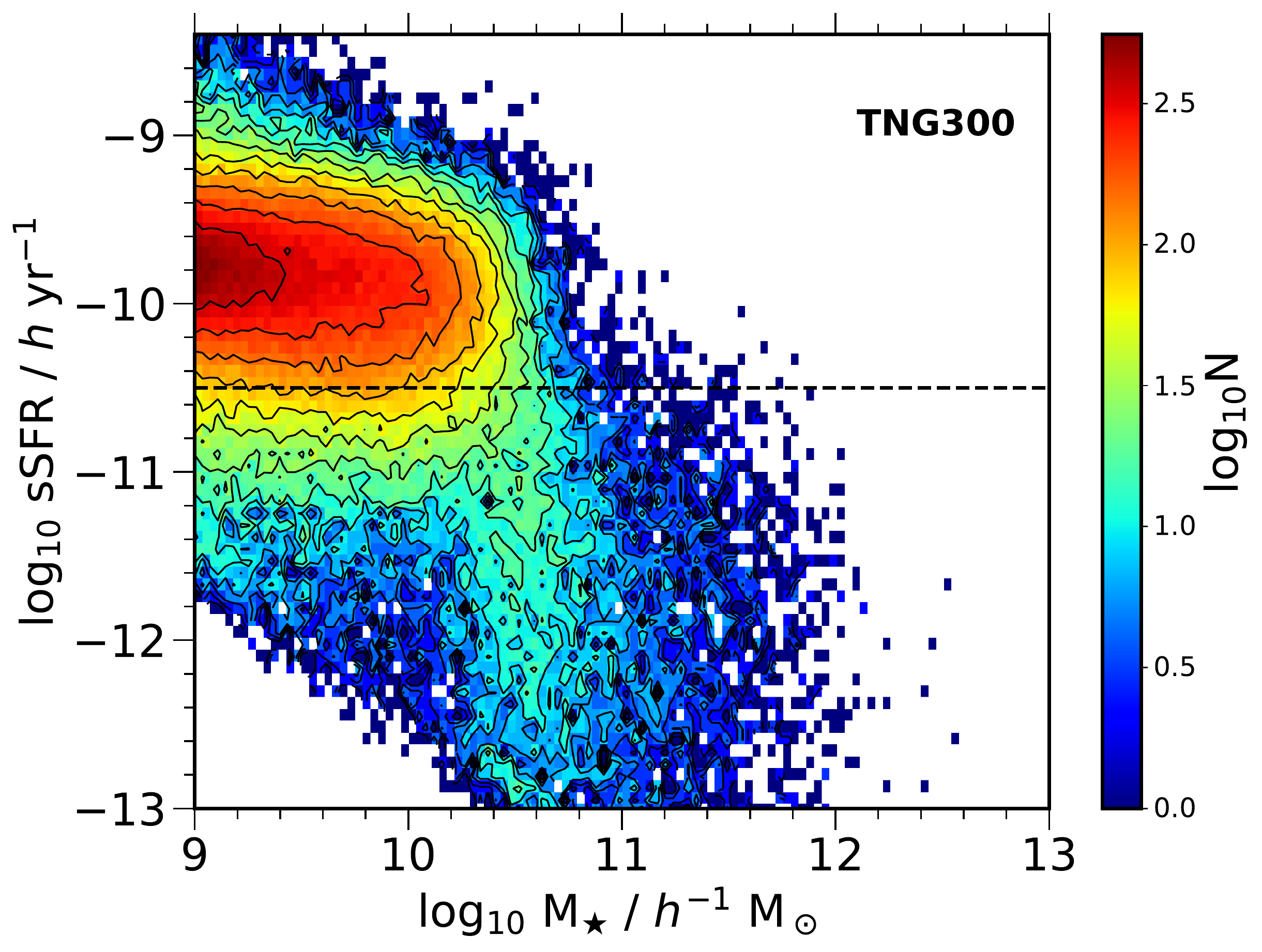}
\includegraphics[width=\columnwidth]{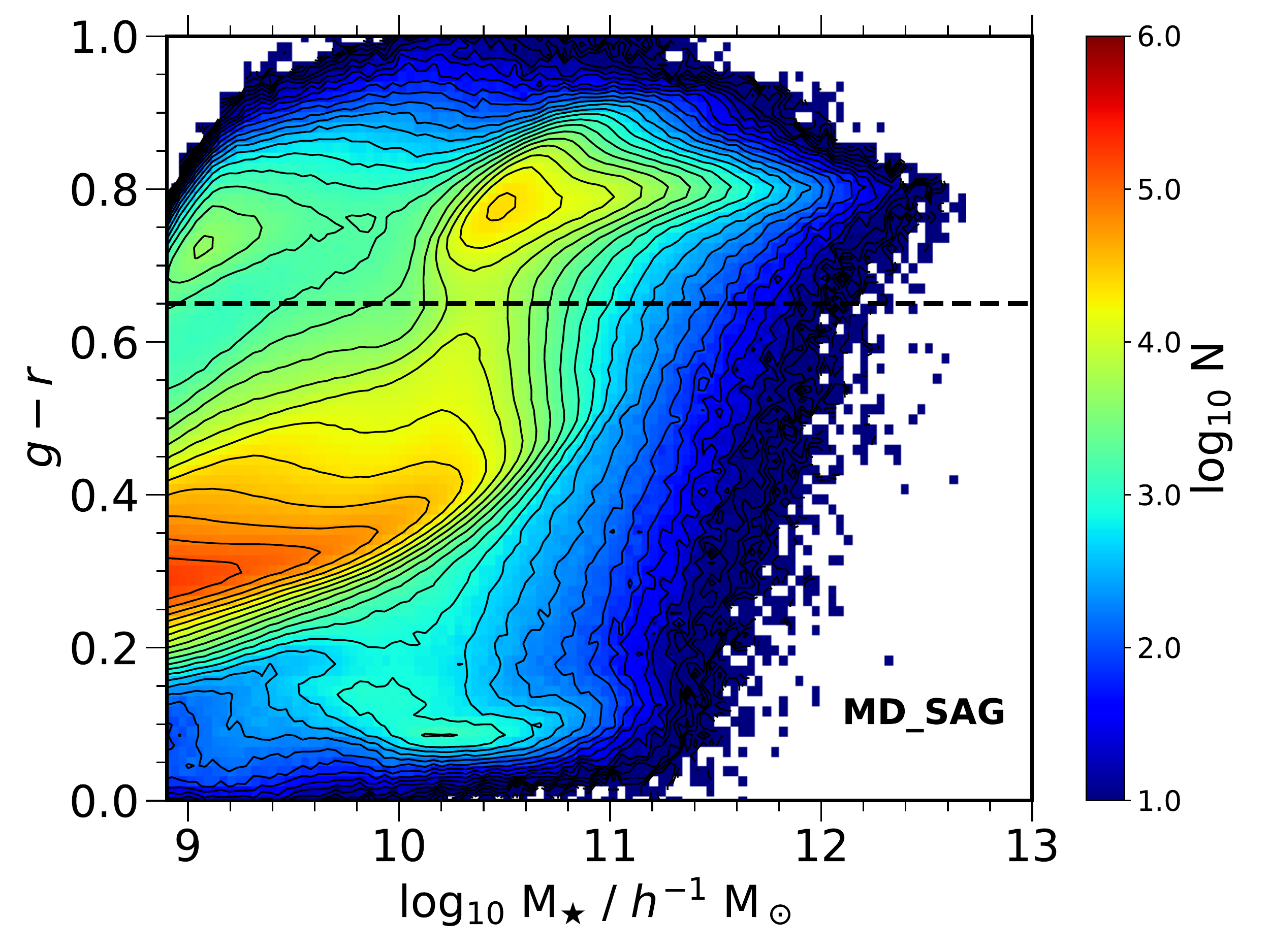}
\includegraphics[width=\columnwidth]{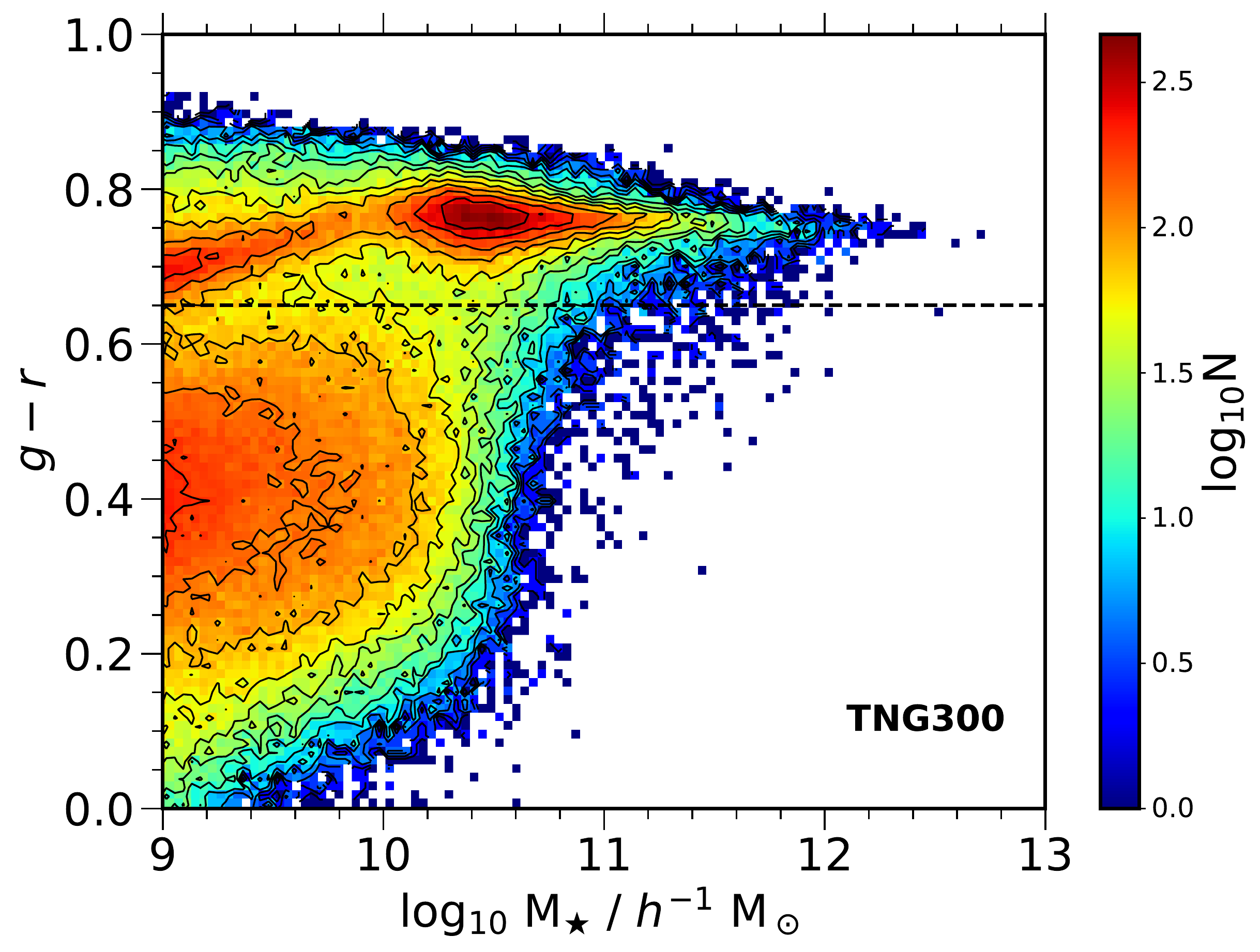}
\caption{
Distributions of sSFR (top panels) and 
$g-r$ colour (bottom panels) as functions of the stellar mass for synthetic galaxies
(centrals and satellites) in the \mdsag~catalogue (left) and \tng{} simulation (right),
represented by contours coloured-coded according to the number density of galaxies, as indicated in the colour bars.
 The black dashed lines depict
 the fiducial values that roughly separates quenched and star-forming galaxies (top) and red and blue galaxies (bottom).  
}
\label{fig_cuts}
\end{figure*}

The \mdsag~galaxy catalogue contains about 370,000 galaxy groups and 
clusters with 
\mh\ $\geq 10^{13}\,\mhalo$,
where \mh\ is the dark matter halo mass within a radius that contains a mean density of 200 times the critical density of the Universe, and about 40 million galaxies 
with stellar mass above
7 $\times$ $10^8$ $\mstar$.
This catalogue includes relevant physical parameters such as the stellar mass, SFR, $ugriz$ magnitudes, bulge-to-total stellar mass ratio, and the distinction between central and satellite galaxies.
The sSFR and $g - r$ colour distributions as functions of the stellar mass are shown in the left column of Fig. \ref{fig_cuts} 
for all galaxies (centrals and satellites) in the \mdsag~catalogue. Similar distributions are obtained for the galaxies in the \mdsagbeta~catalogue (not shown). Although the fraction of quenched galaxies is in better agreement with observational measurements in the latter case, as already mentioned, we decided to show the results of our analysis for both the \mdsag~and \mdsagbeta~catalogues since the former is publicly available\footnote{The catalogue is accessible from the CosmoSim database http://www.cosmosim.org/} \citep{Knebe+2018} and allows the reproducibility of our results.

\subsection{The \tng~hydrodynamical simulation}
\label{sec_TNG}

Cosmological hydrodynamical simulations have the benefit of providing predictions that are less model-dependent than those coming from SAMs, because they follow the evolution of dark matter particles, gas cells, and stellar particles simultaneously in a self-consistent way. Therefore,
we also use the {\sc IllustrisTNG} simulation \citep[][]{Naiman+2018, Nelson+2018, Marinacci+2018, Pillepich+2018_473, Springel+2018, Nelson+2019}.
Among the different boxes, we choose {\sc IllustrisTNG300}, which is one of the largest magneto-hydrodynamical cosmological simulations available, with 
a side length of 205 $\mpc$. 
The \tng{} simulation adopts the standard $\Lambda$CDM cosmology \citep{Planck2016}, with parameters $\Omega_{\rm m} = 0.3089$,  $\Omega_{\rm b} = 0.0486$, $\Omega_\Lambda = 0.6911$, $H_0 = 100\,h\, {\rm km\, s^{-1}Mpc^{-1}}$ with $h=0.6774$, $\sigma_8 = 0.8159$, and $n_s = 0.9667$. It follows the evolution of 2500$^3$ dark-matter particles of mass 4.0 $\times$ 10$^7$ $\mhalo$, and 2500$^3$ gas cells with a mass of 7.6 $\times$ 10$^6$ $\mhalo$. {\sc IllustrisTNG300} contains about 4000 galaxy groups and galaxy clusters with masses between 10$^{13}$ $\leq$ \mh/$\msun$ 
$\leq$ 10$^{15}$, and 
280 galaxy clusters with \mh\ $>$ 10$^{14}$ $\msun$
\citep[][]{Pillepich+2018_475}.  

 The {\sc IllustrisTNG} simulation suite was built using the {\sc arepo} moving-mesh code \citep{Springel2010} and is regarded as an improved version of its predecessor, the {\sc Illustris} simulation \citep{Vogelsberger2014a, Vogelsberger2014b, Genel2014}. Its updated sub-grid models account for a variety of physical precesses, including star formation, radiative metal cooling, chemical enrichment from Type II and Type Ia supernovae (SNe) events, and asymptotic giant branch (AGB) stars, stellar feedback, and super-massive black hole feedback \citep[see][for more details]{Weinberger+2017,Pillepich+2018_473}. Importantly, these models were specifically calibrated to reproduce important observational constraints such as the observed $z=0$ galaxy stellar mass function,\footnote{Note that we have used the standard {\sc IllustrisTNG300} galaxy catalogue. The best agreement between the model and the observational data, as far as the stellar content of haloes is concerned, is obtained with a modified catalogue where stellar masses are slightly re-scaled (see \citealt{Pillepich+2018_475} for more information).} the cosmic SFR density, the halo gas fraction, the galaxy stellar size distributions, or the black hole -- galaxy mass relation.

As for dark matter haloes, these objects are defined in {\sc IllustrisTNG} using a friends-of-friends (FOF) algorithm with a linking length of 0.2 times the mean inter-particle separation \citep{Davis1985}. The gravitationally bound substructures that we call subhaloes are, in turn, identified using the {\sc subfind} algorithm \citep{Springel2001,Dolag2009}. In IllustrisTNG, all subhaloes containing a non-zero stellar mass component are labelled galaxies, but here we use a stellar-mass threshold of 
$10^9$ $\mstar$ 
(see Sec. \ref{sec_method}).
 
The IllustrisTNG simulation has been an excellent tool for studying the connection between galaxies and dark matter haloes at small and large scales.
Among others, it has been implemented to study the occupancy variations \citep{Bose2019,Hadzhiyska2020}, the impact of secondary halo properties on the galaxy population \citep[][]{Montero-Dorta2020,Contreras+2021,MonteroDorta+2021,Favole+2022}, and the galaxy size relation of satellite and central galaxies with their host dark matter haloes \citep{Rodriguez2021}.
The distributions of the sSFR and $g-r$ colour as functions of the stellar mass in the IllustrisTNG300 simulation are shown in the right panels of Fig. \ref{fig_cuts}. 
%

\section{Methodology}
\label{sec_method}

There are different ways to measure correlations
of galaxy properties
between central and neighbouring galaxies. For example, \citet{Kauffmann+2013} measured the sSFR of neighbouring galaxies around isolated galaxies, as observational proxies of central galaxies, as a function of the projected distance from the central galaxies at a given stellar mass bin. This approach has also been used in mock catalogues \citep[e.g.][]{Sin+2017, Tinker+2018}. Another approach is to estimate the mean red fraction or mean quenched fraction of neighbouring (secondary) galaxies as a function of the distance from the central (primary) galaxies in stellar mass bins of the centrals. In simulations, the distance is estimated in real space \citep[e.g.][]{Bray+2016, Lacerna+2018MNRAS}. Here, we use the latter approach %
because the comparison with observations is outside the scope of the current paper. This aspect will 
be a matter of another work. 
Instead, we focus on the results obtained directly from the boxes of the simulations at $z \sim 0$. In this way, we can assess
the results under ideal conditions in the simulations. 

Galaxies with stellar mass \ms\ $>$ $7 \times 10^8$ $\mstar$
in the samples built from the
\sag~model and 
\ms\ $>$ $10^9$ $\mstar$ from
the {\sc IllustrisTNG300} simulation
are classified according to sSFR, $g-r$ colour, stellar mass bins, and halo mass (\mh) bins.
Synthetic galaxies are well modelled down to these thresholds in stellar mass in these catalogues. For instance, the low-mass end of the observed stellar mass function at $z \sim 0$ is well reproduced until this lower limit in the \sag~model \citep[see fig. 1 of][]{Cora+2018}.
For {\sc IllustrisTNG300}, the lower limit in stellar mass corresponds to approximately 
130 gas cells. This threshold is consistent with previous works \citep[e.g.][]{Pillepich+2018_475, Donnari+2019, Montero-Dorta2020, Donnari+2021}.
Furthermore, there is an overall good agreement in these galaxy catalogues with the observed quenched fraction of low-mass central galaxies at low redshift \citep[e.g. see][]{Xie+2020, Donnari+2021}. 
Although these galaxy catalogues tend to overestimate the fraction of low-mass quenched satellite galaxies in low-mass haloes \citep{Cora+2018, Xie+2020, Donnari+2021}, these quenched fractions in simulations reduce when similar conditions to the observations are applied \citep{Donnari+2021}. 

By establishing a cut or threshold in sSFR and colour, we can select the galaxies with quenched star formation and red colours. 
We refer to a galaxy as quenched if the 
${\rm sSFR} \leq 10^{-10.5}$ $h$ yr$^{-1}$,
whereas galaxies 
with sSFR above this value are considered as star-forming galaxies (e.g., see the dashed line in the top panels of Fig. \ref{fig_cuts}). 
The chosen value in sSFR is based on \citet{Brown+2017}
for selecting star-forming galaxies. \citet{Cora+2018} found that this cut in sSFR (10$^{-10.7}$ yr$^{-1}$ with $h=0.678$) allows a better separation between star-forming and quiescent galaxies in the \sag{} model
than other cuts commonly used in the literature, e.g. sSFR = 10$^{-11}$ yr$^{-1}$ \citep{Wetzel+2012}.
We opted for using the same cut in sSFR for the {\sc IllustrisTNG300} galaxy catalogue for consistency, avoiding biased results for particular cuts in each catalogue. 
We obtain a 
fraction of quenched galaxies
of 14 per cent for the \mdsag~galaxy catalogue, 17 per cent for the 
\mdsagbeta~one,
and 38 per cent for the {\sc IllustrisTNG300} simulation. 
Although the fraction of quenched galaxies is smaller in the \sag{} galaxy catalogues compared to the \tng{} one, the top panels of Fig. \ref{fig_cuts} show that both \mdsag{} and {\sc IllustrisTNG300} models have a similar concentration of star-forming galaxies using this cut in sSFR.

Likewise, we refer to a galaxy as red if 
$g-r \ge 0.65$,
whereas galaxies with colours
below this value are considered blue galaxies (e.g. see the dashed line in the bottom panels of Fig. \ref{fig_cuts}).
We selected this colour cut from the bimodality in the \mdsag~galaxy catalogue (bottom left panel in Fig. \ref{fig_cuts}). 
The chosen value is in rough agreement with the separation between the red sequence and the blue cloud in the Sloan Digital Sky Survey (SDSS) galaxies \citep[e.g.][]{BlantonMoustakas2009}.
We obtain a fraction of red galaxies of 12 per cent for the \mdsag{} catalogue.
Again, we opted to use the same colour cut for the other galaxy catalogues for consistency. 
The fraction of red galaxies is 15 per cent for \mdsagbeta{} and 31 per cent for \tng{} catalogues. 
Although the fractions of red galaxies are different between the \sag{} galaxy catalogues and the \tng{} one,
their colour distributions 
are similar, in general. They show a blue cloud and the red sequence, separated by the chosen colour cut.

With the distinction of central and satellite galaxies 
as defined by the corresponding halo finders of each
available catalogue,
we can measure the mean quenched fraction $f_{\rm Q}$
or the mean red fraction $f_{\rm r}$ of neighbouring (secondary) galaxies around central (primary) galaxies at fixed mass to assess the galactic conformity. 
The neighbouring galaxies are all the galaxies (centrals and satellites) above the \ms\ threshold
out to a real-space distance of 10 $\mpc$ from the central galaxies in each sample of primary galaxies.
To measure the galactic conformity at a given distance, we estimate the mean fractions of quenched (red) neighbouring galaxies around both quenched (red) and star-forming (blue) primary galaxies at fixed stellar or halo mass,\footnote{We will estimate the mean fractions of neighbouring galaxies for separations between 0.6 $\mpc$ out to 10 $\mpc$ from the primary galaxies, but we will discuss the results for scales $r$ $\gtrsim$ 1 $\mpc$ because we are interested in the two-halo conformity. At smaller scales, the conformity signal might be mixed with that from the one-halo conformity.}
and calculate the difference between these fractions, 
$\Delta f_{\rm Q}$ ($\Delta f_{\rm r}$).
If the difference is close to zero, there is no correlation between the 
sSFR or colour 
of central galaxies and their neighbouring galaxies, i.e., there is no conformity. 
Therefore, galactic conformity becomes more prominent as the difference, 
$\Delta f_{\rm Q}$, increases.
Although our galaxy catalogues are imperfect in reproducing all the observed trends regarding the absolute quenched fractions, these discrepancies do not significantly affect the differences in the quenched fractions at fixed mass, i.e., the galactic conformity signal.

We will refer to the primary sample with \textit{all} the central galaxies at fixed mass as `PrimAll'. The conformity signal with the `PrimAll' sample is the fiducial case of two-halo conformity measured in simulations and observations.
We then repeat the same procedure, but removing central galaxies in the vicinity around massive systems of \mh\ $\geq$ 10$^{13}$ $\mhalo$
out to 5 $\mpc$.
We choose this scale as a simple representation of the large-scale environment beyond the virial radius of host haloes \citep[e.g.][]{AF+2018}, which is also
a scale that is typically 
larger than filament thickness
\citep{Kuutma+2017}. We discuss the chosen vicinity radius in Sec. \ref{sec_vicinity_radius}. The
central galaxies within the vicinity radius around massive systems
are only removed from the primary sample. 
The secondary sample of neighbouring galaxies out to 10 $\mpc$ of each remaining primary galaxy is 
the same.
We will refer to this primary sample 
that does not include central galaxies around clusters (AC) or groups as \mbox{`PrimNotAC'}.

The analysis of samples \mbox{`PrimAll'} and \mbox{`PrimNotAC'} will allow us to establish the contribution
of central galaxies located around massive systems on the two-halo conformity signal measured at separations of several Mpc.
An excess of conformity signal at few Mpc scales by considering
\textit{all} the central galaxies in `PrimAll'
with respect to the signal corresponding to
the case
`PrimNotAC' will confirm that the two-halo conformity, i.e., correlations of star formation or colour between central galaxies and neighbouring galaxies beyond the virial radius of virialized structures, can be explained mainly by the presence of central galaxies in the outskirts of rich galaxy groups and clusters.

The errors in the estimation of the mean fractions for the \sag{} catalogues are calculated using the jackknife method 
\citep[e.g.][]{Zehavi+2002, Norberg+2009}.
For this, we split every sample into 120 subsamples. Error bars in the mean fractions are estimated using the diagonal
of the covariance matrix. 
Given the large number of galaxies, the error bars are small enough to be 
imperceptible in some of the following figures.

The methodology for estimating errors is slightly different for \tng, but it is also based on a jackknife technique. The box is divided in 8 subboxes ($ L_{{\rm sub-box}} = L_{{\rm box}}/2 = 102.5$ $h^{-1}$Mpc) so that one subbox at a time is disregarded. The uncertainties on the mean fractions correspond to the standard deviation computed from the 8 subvolumes ($V_i = \frac{7}{8} V_{\rm total}$).

\section{Results}
\label{sec_results}

\begin{figure}
\includegraphics[width=\columnwidth]{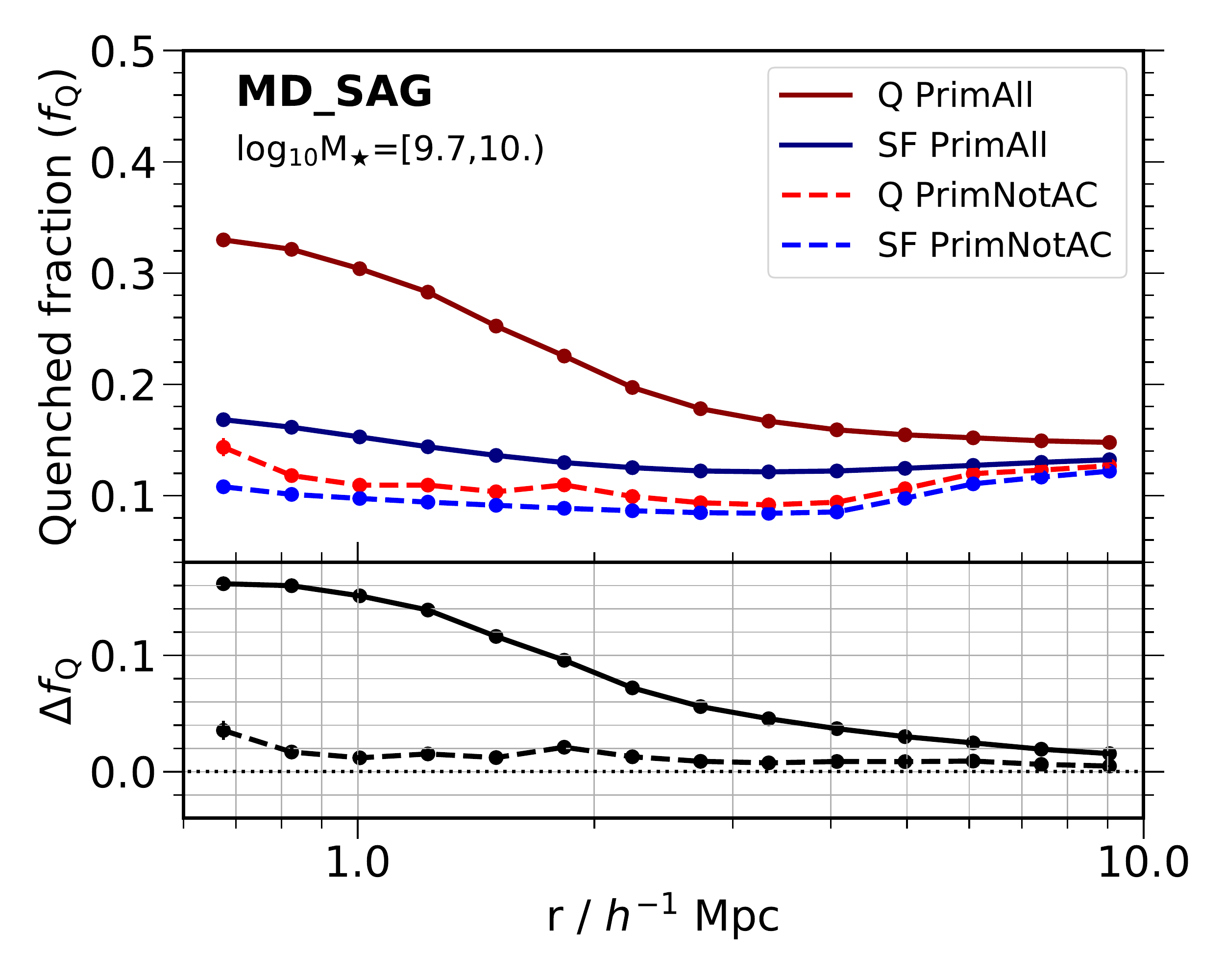}
\includegraphics[width=\columnwidth]{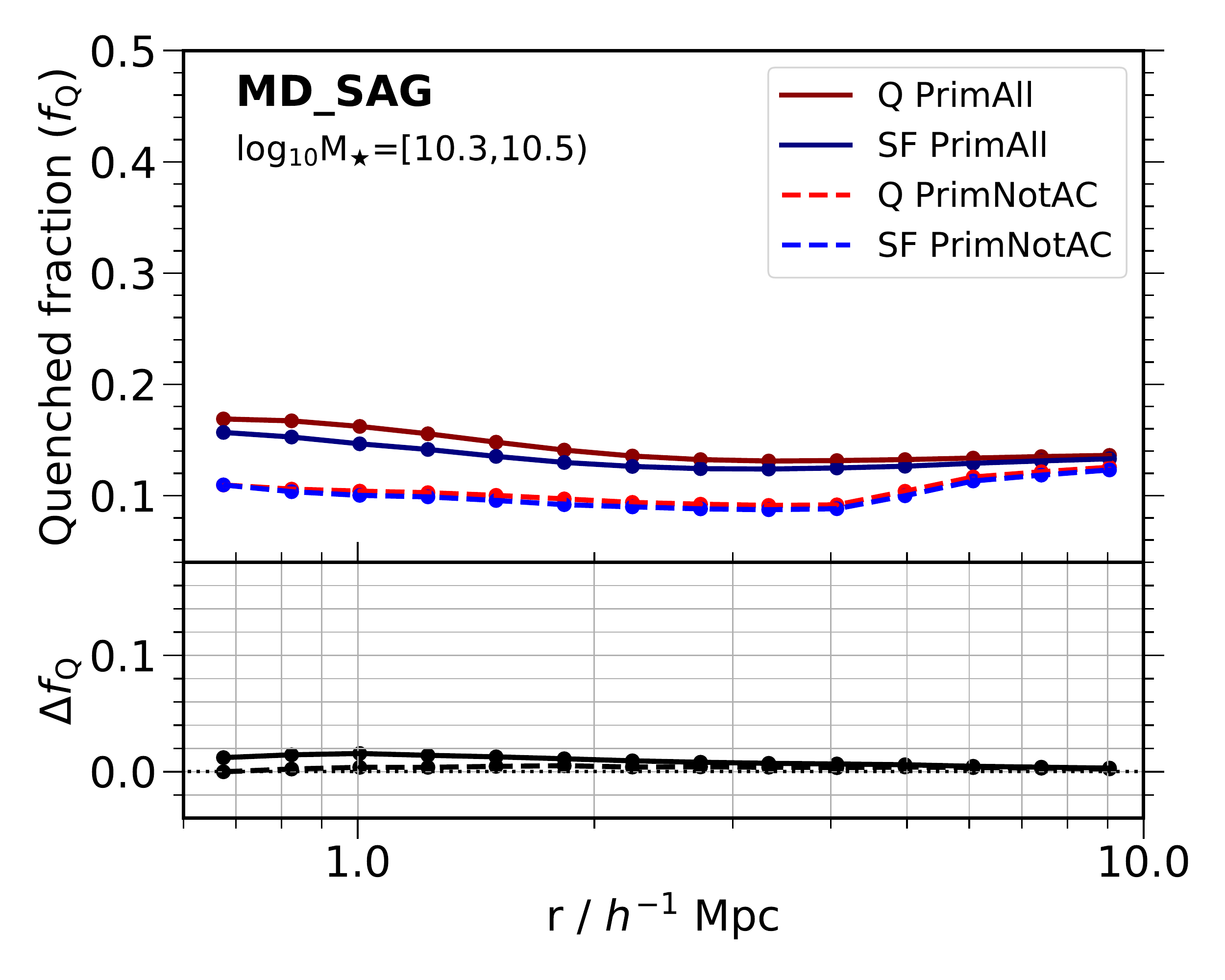}
\caption{Main panels: mean quenched fractions of neighbouring galaxies as functions of the (real-space) distance from the primary galaxies in the 
\mdsag~catalogue.
The primary galaxies are separated
in two stellar-mass bins: 10$^{9.7}$ $\leq$ \ms/$\mstar$ $<$ 10$^{10}$ (top) and 10$^{10.3}$ $\leq$ \ms/$\mstar$ $<$ 10$^{10.5}$ (bottom). 
The solid lines consider all the central galaxies in the primary sample, `PrimAll' (dark red and navy blue for quenched and star-forming primary galaxies).
The dashed lines correspond to the mean fractions after removing the central galaxies in the vicinity of haloes more massive than 10$^{13}$ $\mhalo$ 
from the primary sample, case \mbox{`PrimNotAC'}
(red and blue for quenched and star-forming primary galaxies).
Sub-panels: difference of the mean quenched fractions of neighbouring galaxies around quenched and star-forming primary galaxies at fixed stellar mass. The x-axis is the distance from the primary galaxies. 
The solid line shows the case \mbox{`PrimAll'}, whereas the dashed line is the result obtained for the case \mbox{`PrimNotAC'}.
The dotted line denotes the case of zero difference,
i.e. no conformity.
}
\label{fig_comp_ssfr}
\end{figure}

Figure \ref{fig_comp_ssfr} shows the mean quenched fractions of neighbouring galaxies as functions of the distance from the primary galaxies in two stellar-mass bins in the 
\mdsag~catalogue.\footnote{We arbitrarily chose those stellar mass bins to remark results for low-mass central galaxies \ms\ $< 10^{10}$ $\mstar$ and for intermediate-mass central galaxies with \ms\ $\sim 10^{10.4}$ $\mstar$. 
We confirmed that the results for central galaxies between these two \ms\ bins correspond to a transition in the conformity signal shown in the two panels of Figure \ref{fig_comp_ssfr}. The latter is also valid for the other galaxy catalogues used in this paper.}
The solid lines correspond to 
\textit{all} the central galaxies in the primary sample, `PrimAll'.
The mean quenched fraction ($f_{\rm Q}$) is higher around quenched central galaxies (dark red solid line) compared with that around star-forming centrals (navy blue solid line) 
up to
3 $\mpc$ from low-mass central galaxies (top panels). 
Both fractions tend to converge to the overall quenched fraction of this galaxy catalogue at large scales of about 10 $\mpc$. 
The sub-panels of Fig. \ref{fig_comp_ssfr} show the difference between both mean quenched fractions ($\Delta f_{\rm Q}$) as a function of the distance from the central galaxies for the same stellar mass bins (solid lines). The difference is as big as $\Delta f_{\rm Q}$ $\sim$ 0.15 at $\sim$ 1 $\mpc$ for low-mass primary galaxies of 10$^{9.7}$ $\leq$ \ms/$\mstar$ $<$ 10$^{10}$, and declines to 
$\Delta f_{\rm Q}$ $\lesssim$ 0.05
at distances $r$ $\gtrsim$ 3 $\mpc$. For primary galaxies of 10$^{10.3}$ $\leq$ \ms/$\mstar$ $<$ 10$^{10.5}$, the difference in the quenched fractions of neighbours is always smaller than 0.02. This result confirms that the two-halo conformity is much stronger for low-mass central galaxies \citep[e.g.][]{Kauffmann+2013, Bray+2016, Lacerna+2018MNRAS}.

For the sample `PrimNotAC', in which the central galaxies in the vicinity of groups and clusters of galaxies are removed from the primary sample, the mean quenched fraction is slightly higher around quenched central galaxies compared with that around star-forming centrals (red and blue dashed lines in Fig. \ref{fig_comp_ssfr}, respectively). 
The mean quenched fractions in `PrimNotAC' increase to converge to the overall fraction of quenched galaxies of this catalogue from distances $\gtrsim$ 5 $\mpc$ because of the spatial condition in which we remove the centrals in the vicinity of massive structures.
The sub-panels of Fig. \ref{fig_comp_ssfr} show 
the differences in the quenched fractions (dashed line).
These differences are much smaller than the case `PrimAll' of the same stellar mass,  
with $\Delta f_Q$ $\lesssim$ 0.02 for scales $r$ $\gtrsim$ 1 $\mpc$ in both stellar mass bins. 
The excess of two-halo conformity at fixed stellar mass in 
`PrimAll'
compared with the case 
\mbox{`PrimNotAC'}
supports the claim that most of the conformity signal at few Mpc scales can mainly be explained by the central galaxies in the outskirts of rich galaxy groups and clusters.

\begin{figure}
\includegraphics[width=\columnwidth]{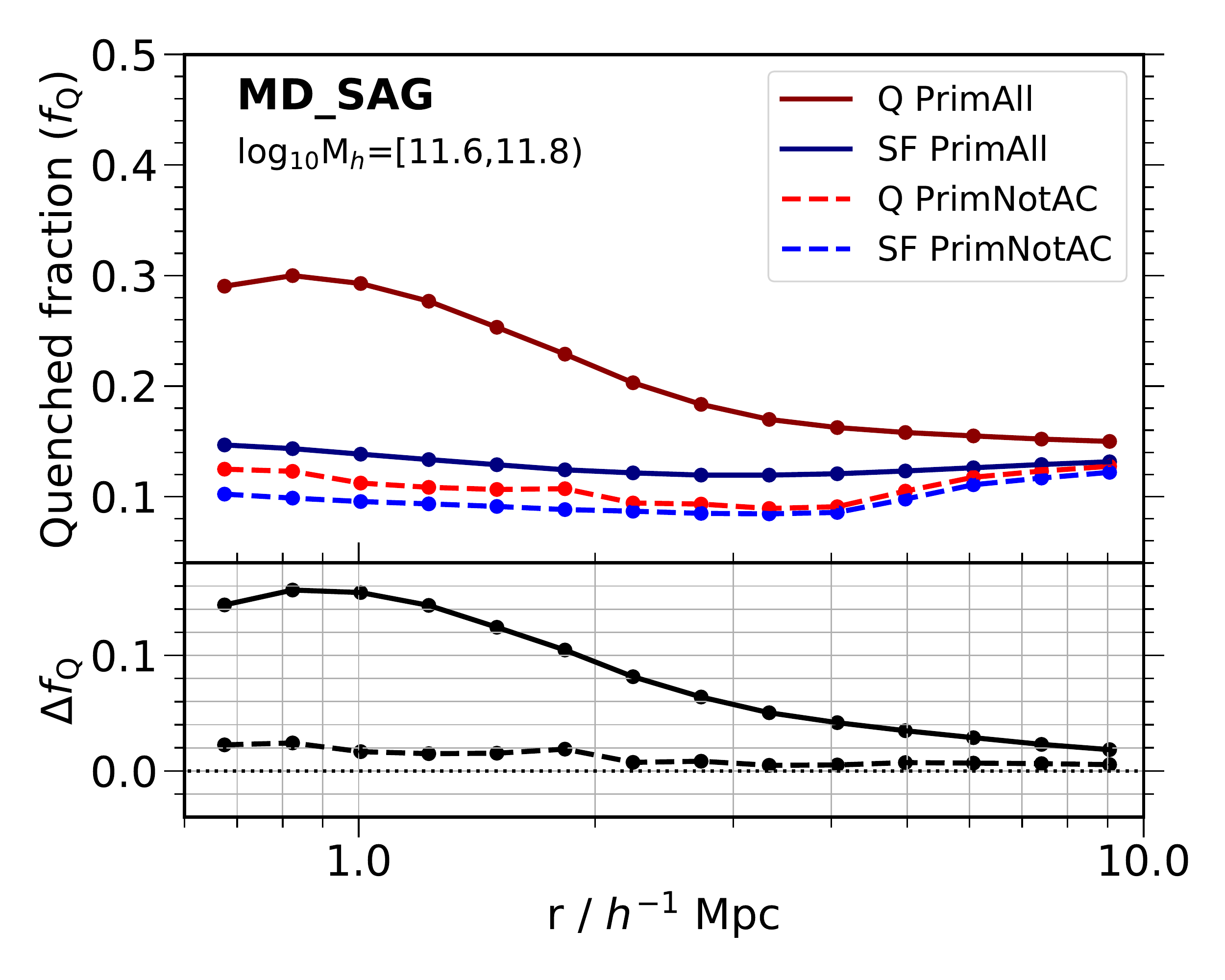}
\includegraphics[width=\columnwidth]{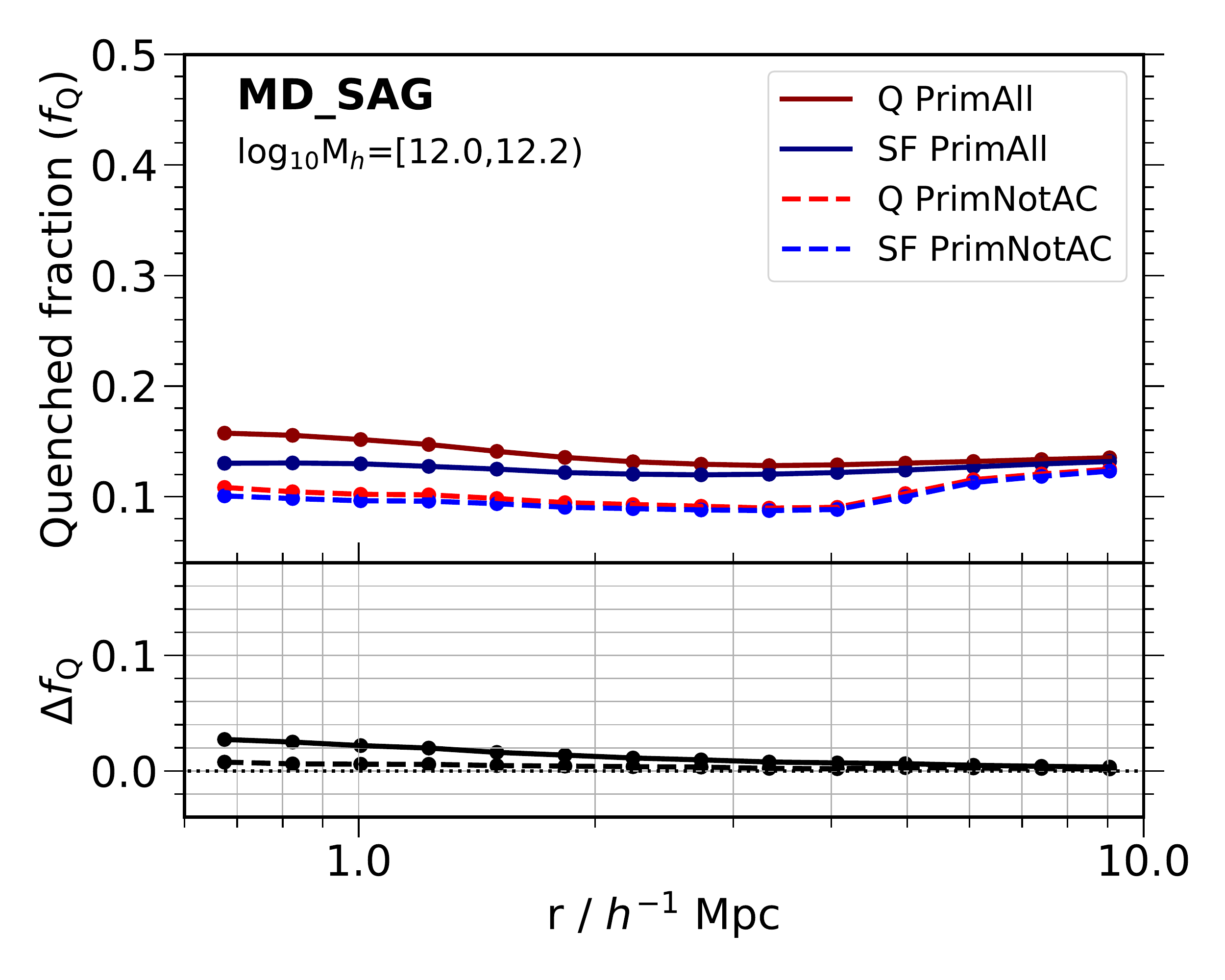}
\caption{Same as Fig. \ref{fig_comp_ssfr} but for primary galaxies at fixed halo mass in the \mdsag~catalogue
(from top to bottom: 
10$^{11.6}$ $\leq$ \mh/$\mhalo$ $<$ 10$^{11.8}$ and 10$^{12}$ $\leq$ \mh/$\mhalo$ $<$ 10$^{12.2}$). 
}
\label{fig_comp_ssfr_fixedMhalo}
\end{figure}

The results above assume 
that the correlations in sSFR between central and neighbouring galaxies
are given by the stellar mass of the primary galaxies, 
as the resulting two-halo conformity is stronger for low-mass primary galaxies.
However, some authors \citep[e.g.][]{Paranjape+2015, Tinker+2018, Treyer+2018} have suggested that the observed conformity at $\lesssim$ 4 Mpc is because red (quenched) 
central galaxies can reside in more massive dark matter haloes than blue centrals of the same stellar mass.
In this regard, for \ms\ $>$ 10$^{10}$ $\mstar$, the median halo mass of quenched central galaxies is higher than that of star-forming central galaxies at fixed stellar mass \citep{Lacerna+2018MNRAS}.
Fig. \ref{fig_comp_ssfr_fixedMhalo} 
shows the same exercise as before but for primary galaxies at fixed halo mass. We find qualitatively the same results as for primary galaxies at fixed stellar mass, i.e., there is an excess of correlation between quenched neighbouring galaxies and quenched primary galaxies out to several Mpc distances in the case 
`PrimAll'
(solid lines) compared with the case 
`PrimNotAC'
(dashed lines). The correlation is stronger for primary galaxies in lower mass haloes, but it is drastically reduced 
for the case in which the central galaxies in the vicinity of galaxy groups and clusters are removed from the primary sample at fixed halo mass. 
For instance, for primary galaxies at 
10$^{11.6}$ $\leq$ \mh/$\mhalo$ $<$ 10$^{11.8}$,
the difference of the mean quenched fractions of neighbours is $\Delta f_{\rm Q}$ $\sim$ 
0.15 (0.1) at $r$ $\sim$ 1 (2)
$\mpc$  from the primary galaxies in the case `PrimAll',
but it is $\Delta f_{\rm Q}$ $\lesssim$ 0.02 
at scales $r$ $\gtrsim$ 1 $\mpc$ from primary galaxies in the case \mbox{`PrimNotAC'} with the same host halo mass. 
For primary galaxies at intermediate halo masses (10$^{12}$ $\leq$ \mh/$\mhalo$ $<$ 10$^{12.2}$, bottom panel), both 
cases `PrimAll' and `PrimNotAC'
show $\Delta f_{\rm Q}$ $\lesssim$ 0.02 at distances $r$ $\gtrsim$ 1 $\mpc$.
Therefore, the conformity measured at few Mpc scales is largely driven by low-mass central galaxies or central galaxies in low-mass haloes located in the outskirts of galaxy groups and galaxy clusters.

\begin{figure}
\includegraphics[width=\columnwidth]{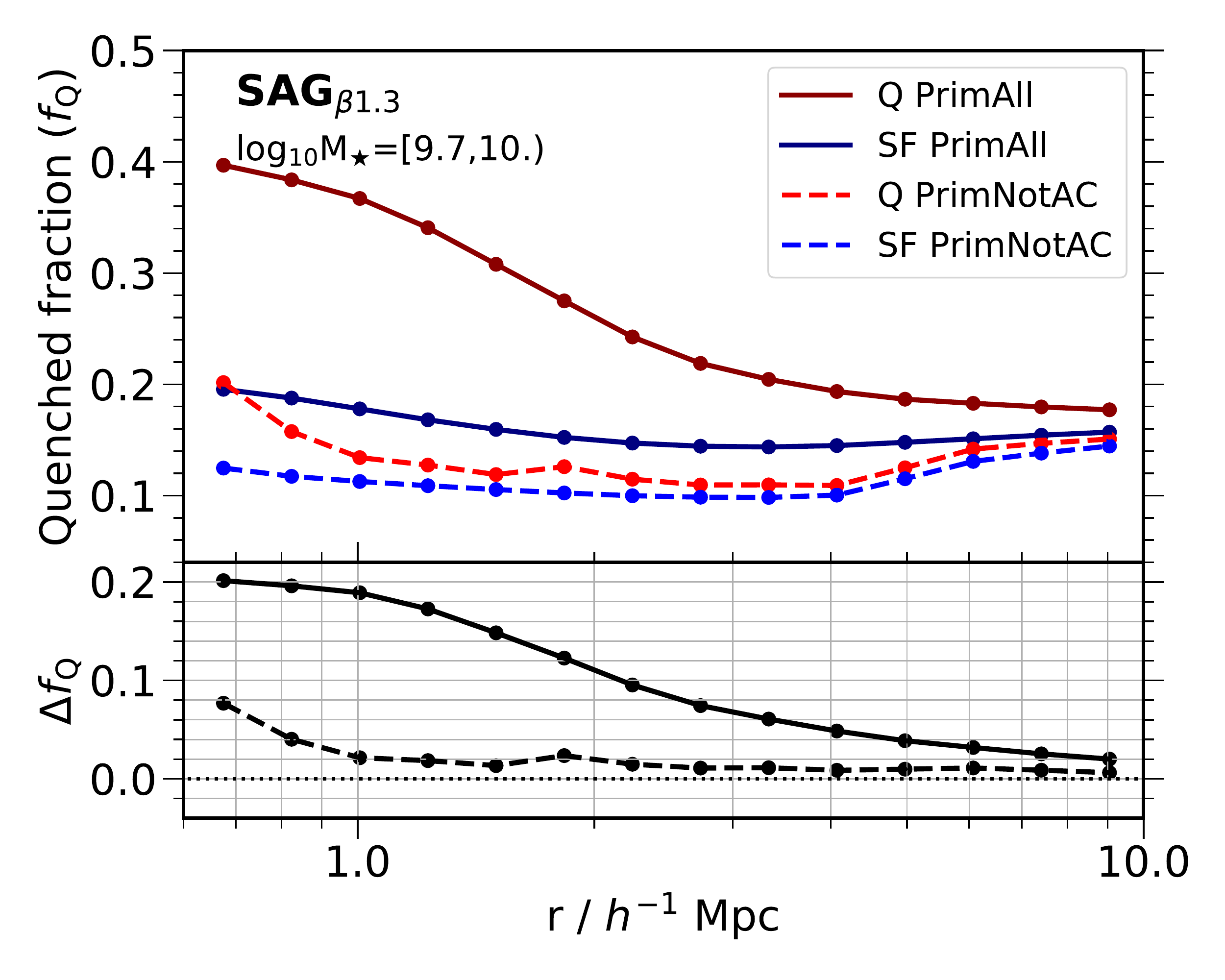}
\includegraphics[width=\columnwidth]{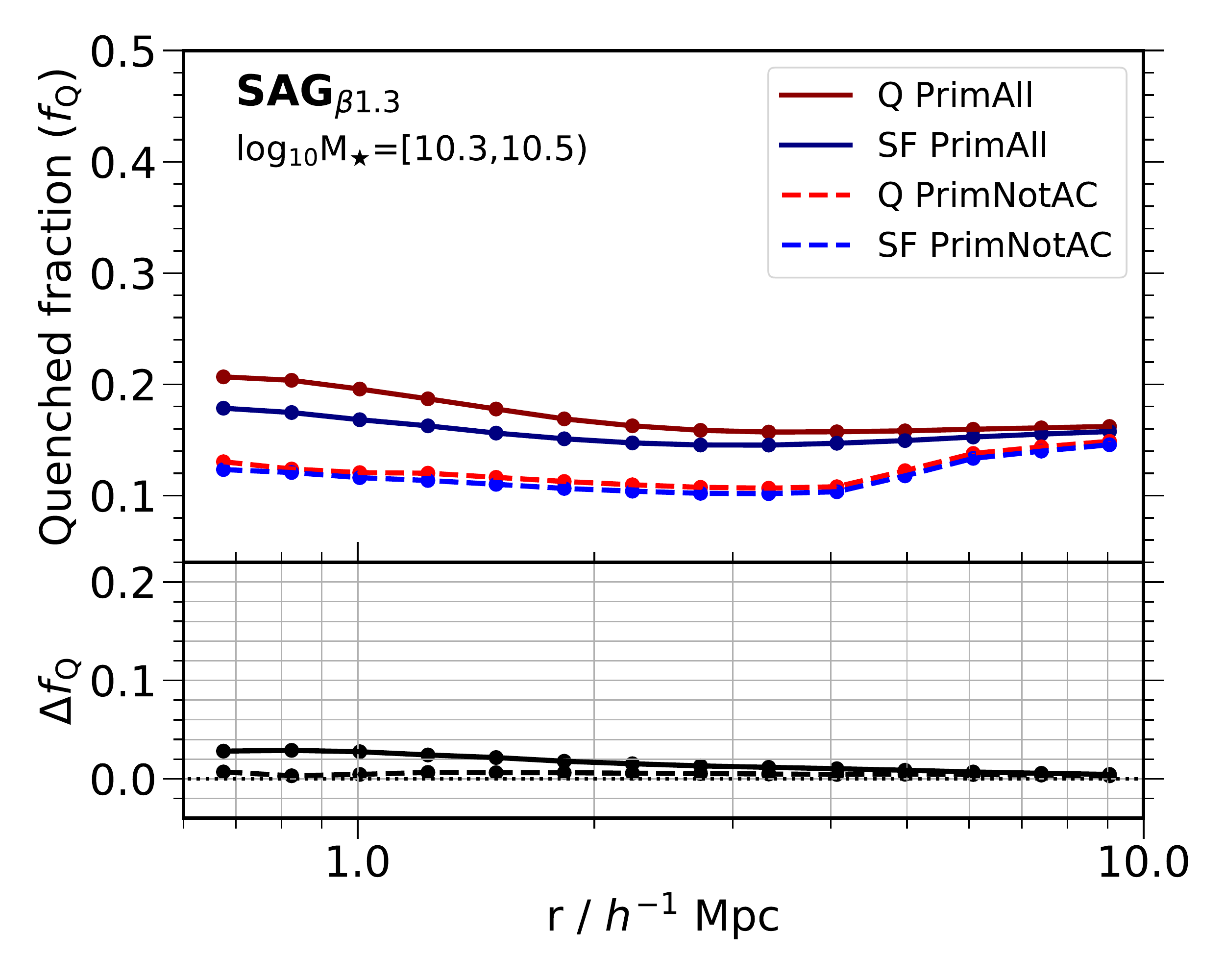}
\caption{Same as Fig. \ref{fig_comp_ssfr} but for primary galaxies at fixed stellar mass in the 
\mdsagbeta~galaxy catalogue.
}
\label{fig_comp_ssfrBeta}
\end{figure}

We now repeat the same test but using the 
\mdsagbeta~galaxy catalogue.
The results at fixed stellar mass for the primary galaxies are shown in Fig. \ref{fig_comp_ssfrBeta}. 
The trends are very similar to those 
obtained from the \mdsag~catalogue
(Fig. \ref{fig_comp_ssfr}).
Here, the conformity reduces almost 
one order of magnitude at $r \sim$ 1 $\mpc$ and a factor of three at $r \sim$ 3 $\mpc$ in the case `PrimNotAC' compared with the case `PrimAll'
for low-mass primary galaxies in the \mdsagbeta{} model.

\begin{figure}
\includegraphics[width=\columnwidth]{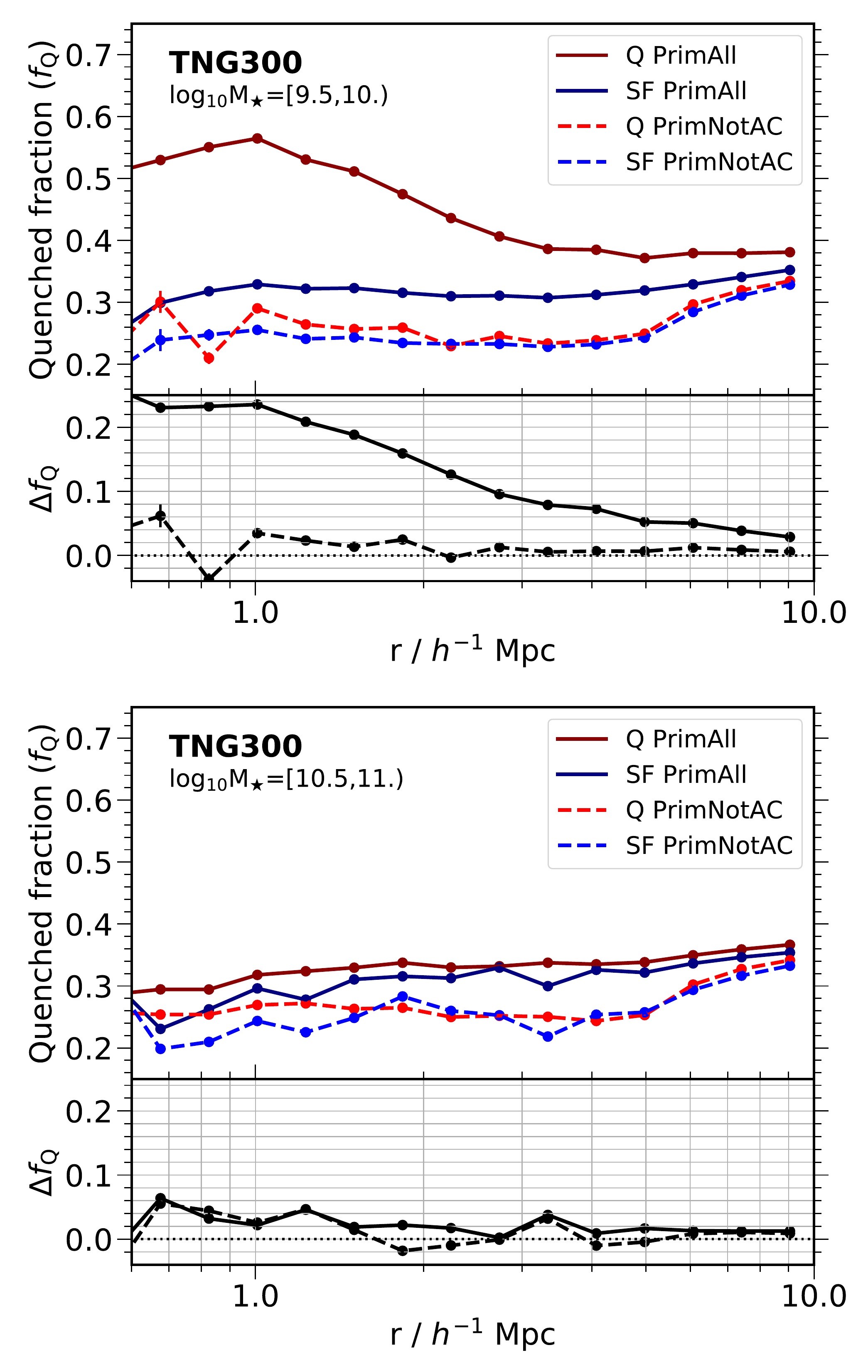}
\caption{Similar to Fig. \ref{fig_comp_ssfr} but for primary galaxies at fixed stellar mass in the \tng{} model (from top to bottom: 10$^{9.5}$ $\leq$ \ms/$\mstar$ $<$ 10$^{10}$ and 10$^{10.5}$ $\leq$ \ms/$\mstar$ $<$ 10$^{11}$). }
\label{fig_fQ_IllustrisTNG}
\end{figure}

The modelling of the physical processes that affect the baryonic components implemented in the \tng{} hydrodynamical simulation differs from the treatment included in the semi-analytic model from which the galaxy catalogues \mdsag~and \mdsagbeta~are built.
Furthermore, the number of synthetic galaxies in \tng{} is smaller compared with the \sag{} catalogues due to the different volume sizes of each simulation.
Therefore, the results 
obtained from \tng{} are shown in bigger stellar-mass ranges.
The case with \textit{all} the central galaxies, ``PrimAll'', is 
shown in Fig. \ref{fig_fQ_IllustrisTNG} as solid lines. For the lowest 
stellar mass bin
10$^{9.5}$ $\leq$ \ms/$\mstar$ $<$ 10$^{10}$ (top panels), there is evident two-halo conformity out to $r \sim$ 5 $\mpc$.
The difference in the mean quenched fractions of neighbours around low-mass quenched and low-mass star-forming primary galaxies is $\Delta f_{\rm Q}\sim $ 0.24 at $r \sim$ 1 $\mpc$ and it reduces
to $\Delta f_{\rm Q}$ $\lesssim$ 0.05 at distances $r \gtrsim$ 5 $\mpc$ from the primary galaxies (black solid line in the sub-panel). 
In contrast, we do not observe a particular correlation with sSFR between the more massive primary galaxies of 10$^{10.5}$ $\leq$ \ms/$\mstar$ $\leq$ 10$^{11}$ and their neighbouring galaxies (bottom panels), where
the difference
is typically $\Delta f_{\rm Q} \lesssim$ 0.04 at distances $\gtrsim$ 1 $\mpc$ from the intermediate-mass primary galaxies (black solid line in the sub-panel).

Figure \ref{fig_fQ_IllustrisTNG}  
also shows the mean quenched fractions of neighbours 
in the case `PrimNotAC' (dashed lines). As in the SAMs, the two-halo conformity is very small or absent in this case. 
For the lowest \ms{} bin, 
10$^{9.5}$ $\leq$ \ms/$\mstar$ $<$ 10$^{10}$, the difference in the mean quenched fractions of neighbours around quenched and star-forming primary galaxies in the case  `PrimNotAC' is typical $\Delta f_Q$ $<$ 0.04 at distances larger than 1 $\mpc$ from the low-mass primary galaxies (black dashed line in the sub-panel). A similar result is obtained for the other 
massive
bin (bottom panels), which resembles the small conformity signal of the fiducial case ``PrimAll'' in this bin.
Therefore, it is a robust result that the conformity measured at few Mpc scales is mainly driven by low-mass central galaxies in the vicinity of groups and clusters because it is independent of the specific model.

In Appendix \ref{sec_colour}, we show the two-halo conformity using the $g - r$ colour. The differences in the mean red fractions in the cases `PrimAll' and `PrimNotAC' are qualitatively very similar to the respective differences in the mean quenched fractions. Therefore, our results are consistent either using sSFR or galaxy colour.

\section{Discussion}
\label{sec_discussion}

\begin{figure}
\includegraphics[width=\columnwidth]{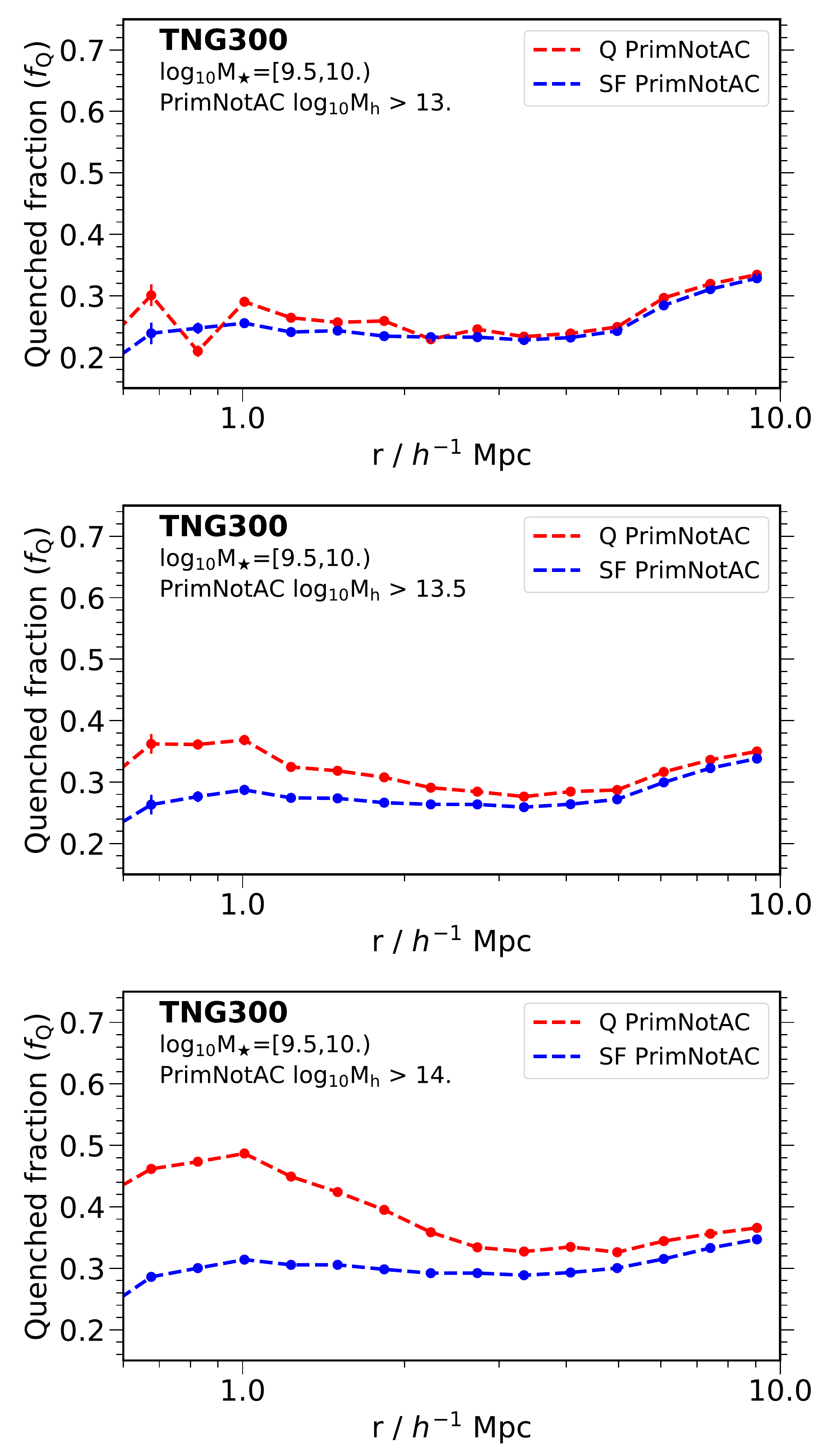}
\caption{Mean quenched fractions of neighbouring galaxies as functions of the distance from the primary galaxies 
with stellar masses in the range
10$^{9.5}$ $\leq$ \ms/$\mstar$ $<$ 10$^{10}$ 
after removing from the primary sample the central galaxies in the vicinity of haloes more massive than 10$^{13}$ $\mhalo$ (top), 10$^{13.5}$ $\mhalo$ (middle), and 10$^{14}$ $\mhalo$ (bottom) in the \tng{} simulation.
The red and blue dashed lines 
represent the values
for quenched and star-forming primary galaxies, respectively.
}
\label{fig_fQnotAC_IllustrisTNG}
\end{figure}

\subsection{Two-halo conformity signal after removing central galaxies}

We have shown in Sect. \ref{sec_results} that the conformity signal at few Mpc decreases drastically when the central galaxies in the vicinity of groups and clusters with \mh\ $\geq$ 10$^{13}$ $\mhalo$ are removed from the primary sample
(`PrimNotAC').
We explore how this signal depends on the halo mass threshold to exclude nearby central galaxies from the primary sample.
Fig. \ref{fig_fQnotAC_IllustrisTNG} shows the mean quenched fractions of neighbours as functions of the distance from low-mass primary galaxies in the \tng{} simulation after removing from the primary sample the central galaxies in the vicinity of haloes of different masses. The top panel shows the results for 
haloes more massive than 10$^{13}$ $\mhalo$, i.e. the lines 
show the same values
as the dashed lines in the top panel of Fig. \ref{fig_fQ_IllustrisTNG}.
The middle and bottom panels of Fig. \ref{fig_fQnotAC_IllustrisTNG} show the results obtained when we remove central galaxies in the vicinity of haloes more massive than 10$^{13.5}$ and 10$^{14}$ $\mhalo$, respectively. The two-halo conformity increases as the 
halo mass threshold
of the systems increases, but the signal at distances $\gtrsim$ 1 $\mpc$ is always smaller than that for the case with \textit{all} the central galaxies in the primary sample (`PrimAll' sample; solid lines in the top panel of Fig. \ref{fig_fQ_IllustrisTNG}). The importance of this result is that
the two-halo conformity is not produced in the outskirts of galaxy clusters only;
we also need to consider the environmental effects in the vicinity of galaxy groups as smaller as 10$^{13}$ $\mhalo$
on low-mass central galaxies 
to largely explain the conformity measured at distances of a few Mpc from the primary galaxies. 
We note we obtain the same results 
from the \mdsag~and \mdsagbeta~catalogues.

\begin{figure}
\includegraphics[width=\columnwidth]{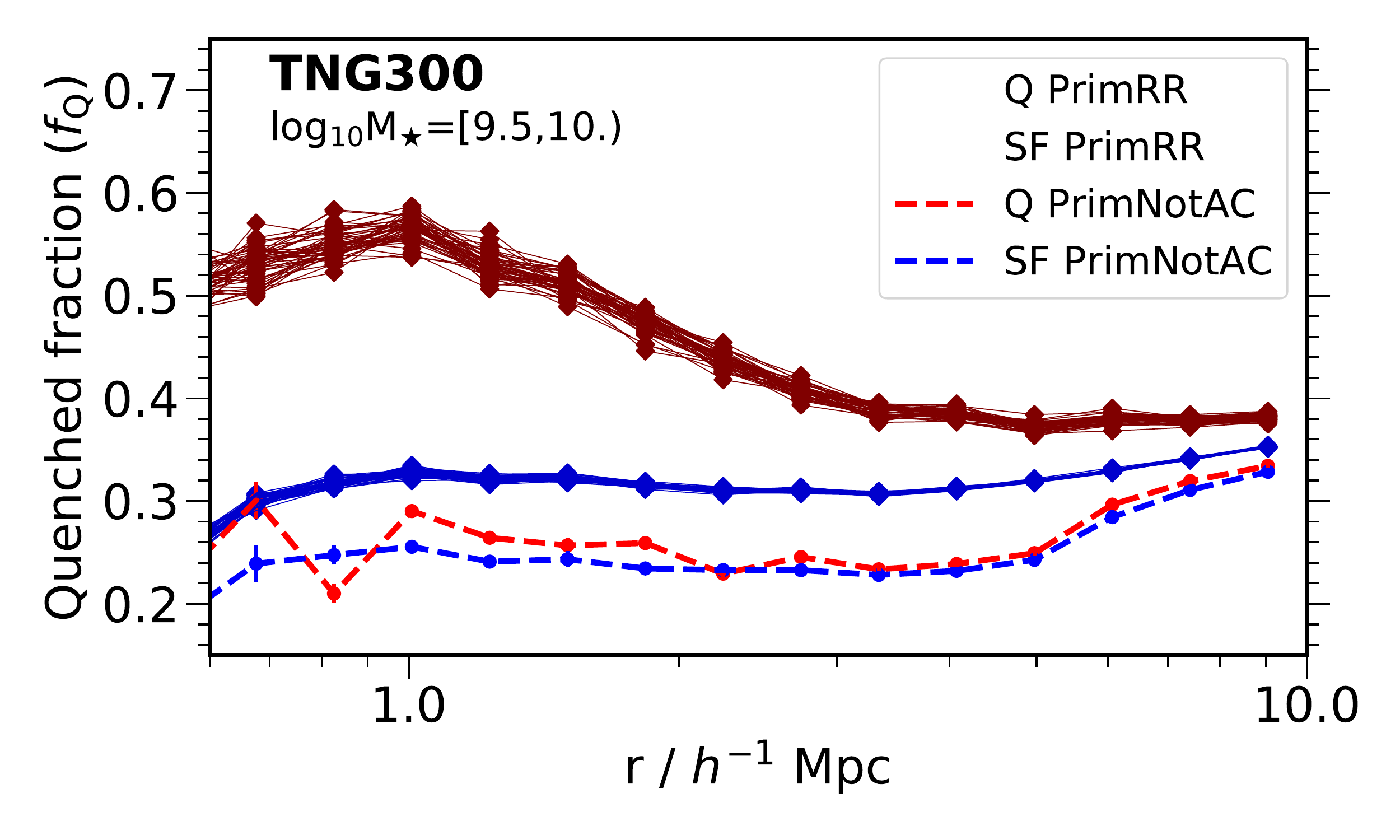}
\caption{Similar to the top main panel of Fig. \ref{fig_fQ_IllustrisTNG}, where the dashed lines correspond to the 
`PrimNotAC' sample,
but we include 50 realizations of randomly removed quenched and star-forming central galaxies from the
`PrimAll'
sample in the \tng{} simulation (maroon and medium blue diamonds, respectively, connected with solid lines). 
}
\label{fig_testA_TNG300}
\end{figure}

We test if the results shown in Sect. \ref{sec_results} are robust or are just an artefact of removing galaxies. 
For this exercise, 
we did 
50 realizations of removing 
randomly
quenched and star-forming central galaxies from the primary sample 
in the \tng{} simulation.
The number of quenched and star-forming central galaxies removed randomly is similar to the case of removing quenched and star-forming central galaxies of the same stellar mass in the vicinity of haloes more massive than 10$^{13}$ $\mhalo$, \mbox{`PrimNotAC'}.
Fig. \ref{fig_testA_TNG300} shows the mean quenched fractions of neighbouring galaxies out to $\sim$ 10 $\mpc$ from low-mass primary galaxies (10$^{9.5}$ $\leq$ \ms/$\mstar$ $<$ 10$^{10}$) for these realizations as solid lines. 
The results are different than the case `PrimNotAC' shown in dashed lines, but they are similar compared with \mbox{`PrimAll'} (see the solid lines in the top main panel of Fig. \ref{fig_fQ_IllustrisTNG}), i.e., the fiducial conformity signal in the sample `PrimAll' remains after randomly removing central galaxies from the primary sample. 
This result supports our claim that most of the galactic conformity measured out to scales of a few Mpc is produced by low-mass central galaxies located near large groups and clusters of galaxies.

We also tested with the \mdsag~catalogue that removing central galaxies randomly from the primary sample does not modify the quenched fractions of neighbours at distances of a few Mpc from the primary galaxies in \mbox{`PrimAll'}.
Therefore, the results shown in the previous section are not an 
artefact of a spurious emergence of trends due to removing, by chance, a highly biased subset of galaxies. 
The results from
both SAMs and the hydrodynamic simulation are consistent,
indicating
that most of the galactic conformity measured at few Mpc scales is produced by low-mass central galaxies in the vicinity of large groups and clusters.

\subsection{Two-halo conformity signal after varying the vicinity radius}
\label{sec_vicinity_radius}

We have considered a radius out to 5 $\mpc$ as the scale of influence of massive systems of \mh\ $\geq$ 10$^{13}$ $\mhalo$, which 
corresponds to a large-scale environment beyond the virial radius of haloes.
Here, we test if the previous results are robust after varying this radius out to
which we remove the central galaxies from the primary sample. 

\begin{figure}
\includegraphics[width=\columnwidth]{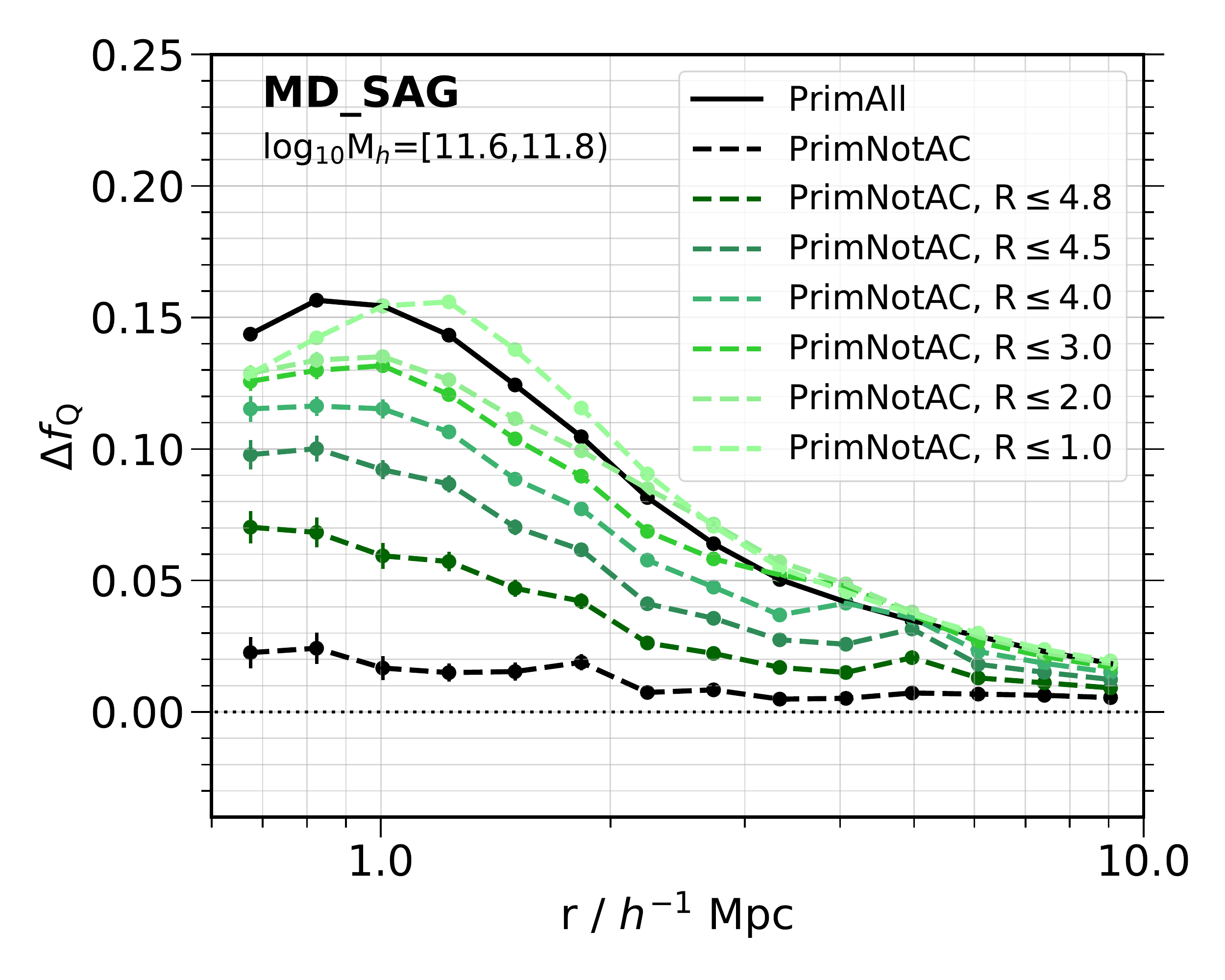}
\caption{Similar to top sub-panel of Fig. \ref{fig_comp_ssfr_fixedMhalo}, but we include the cases of removing from the primary sample the central galaxies in the vicinity of haloes more massive than 10$^{13}$ $\mhalo$ out to 1, 2, 3, 4, 4.5, and 4.8 $\mpc$ (coloured dashed lines according to the legend).
}
\label{fig_diff_Mh_R}
\end{figure}

We test
different vicinity radii starting from 1 $\mpc$. We remove from the primary sample the central galaxies inside the vicinity radius of massive systems of \mh\ $\geq$ 10$^{13}$ $\mhalo$ and estimate the mean quenched fractions of neighbouring galaxies around the (remaining) primary galaxies at fixed halo mass. The results for primary galaxies in host dark matter haloes of
10$^{11.6}$ $\leq$ \mh/$\mhalo$ $<$ 10$^{11.8}$ 
are shown in Fig. \ref{fig_diff_Mh_R}. 
We chose this halo mass range because the fiducial case `PrimAll' shows an evident two-halo conformity signal (see top panel of Fig. \ref{fig_comp_ssfr_fixedMhalo}).
The difference in the mean quenched fractions of neighbouring galaxies, i.e., the conformity signal  $\Delta f_{\rm Q}(r)$, typically decreases as the vicinity radius increases. For example, at a distance of $r$ $\sim$ 3 $\mpc$ from the primary galaxies, the conformity signal $\Delta f_{\rm Q}(3)$ is nearly the same as for the case 
`PrimAll' (black solid line) and the case of removing from the primary sample the central galaxies in the vicinity 
of haloes more massive than 10$^{13}$ $\mhalo$ 
within 1 and 3 $\mpc$, 
with $\Delta f_{\rm Q}(3)\sim 0.06$.
The conformity signal reduces down to $\Delta f_{\rm Q}(3)$ $\sim$ 0.04 
for the case of removing from the primary sample the central galaxies in the vicinity 
of relatively massive haloes 
within 4 $\mpc$. 
The conformity signal is $\Delta f_{\rm Q}(3)$ $\lesssim$ 0.03 after removing from the primary sample the central galaxies in the vicinity 
of haloes more massive than 10$^{13}$ $\mhalo$ within 4.5 and 5 $\mpc$. 

Therefore, the conformity signal at few Mpc scales depends on the vicinity radius around relatively massive haloes. This result can be explained 
as a combination of the radial influence of haloes 
that decreases with the clustercentric radius \citep[e.g.,][]{Wetzel+2012}
and that the exact vicinity radius
depends on the mass, 
i.e., it is smaller for less massive groups and larger for more massive clusters \citep[e.g.,][]{Bahe+2013}.
The choice of a relatively large vicinity radius of 5 $\mpc$ as the scale of the environmental influence of massive systems on nearby central galaxies is probably considering most of the relevant scales in which this effect has been acting. 
\cite{Zinger+2018} find the star formation quenching via `starvation' can occur in the outskirts of simulated galaxy clusters because their hot X-ray emitting intracluster medium can extend out to $\sim$ 2--3 virial radii, which corresponds to median scales of 4--6 Mpc in their sample of clusters, in agreement with our scale of 5 $\mpc$.
Similarly, \cite{Ayromlou+2021} find the quenched fraction of galaxies in the vicinity of groups and clusters is higher than in the field out to $\sim$ 2--3 virial radii using a SAM.
We have tested that $\Delta f_{\rm Q}(r)$ values at distances $r$ $\gtrsim$ 1 $\mpc$ from the primary galaxies are similar for vicinity radii between 5 and 7 $\mpc$, indicating that the environmental influence of massive systems on low-mass central galaxies would not reach scales much more extensive than 5 $\mpc$ in our galaxy catalogues.
An exploratory analysis in which the vicinity radius varies with the halo mass or other halo properties of the groups and clusters will be presented elsewhere.

\subsection{Central galaxies around overdensities}
\label{Sec_AOC}

We have tested 
results from the `PrimNotAC' sample, built by removing
central galaxies in the vicinity of groups and clusters from the primary sample. We have considered the halo mass to identify the central galaxies around massive systems.  Observationally, it is not easy to infer the halo mass because one has to rely on group finder methods that may contain systematics in the halo mass estimation \citep[e.g.,][]{Calderon+2018, Tinker2020arXiv200712200}. 

Here, we explore an alternative approach of using dense environments (overdensities) instead of massive haloes. Overdense systems have the advantage that they can be more easily identified in observations depending on the environmental definition. We use the $\Sigma_N$ parameter, which is a simple estimator to measure the galaxy environment in observations 
\citep[e.g.,][]{Dressler1980, Aguerri2009, Dominguez2002,Nigoche-Netro2019}, 
with the definition given in \citet{Nigoche-Netro2019}, i.e., $\Sigma_N = N/(\pi d_{N}^2)$, 
where $d_N$ is the (real-space) distance to the $N$th nearest neighbouring galaxy. 
We choose $N = 5$ to measure $\Sigma_5$ for each central galaxy using all the galaxies above $7 \times 10^8$ $\mstar$ in the 
\mdsagbeta~galaxy catalogue.

\begin{figure}
\includegraphics[width=\columnwidth]{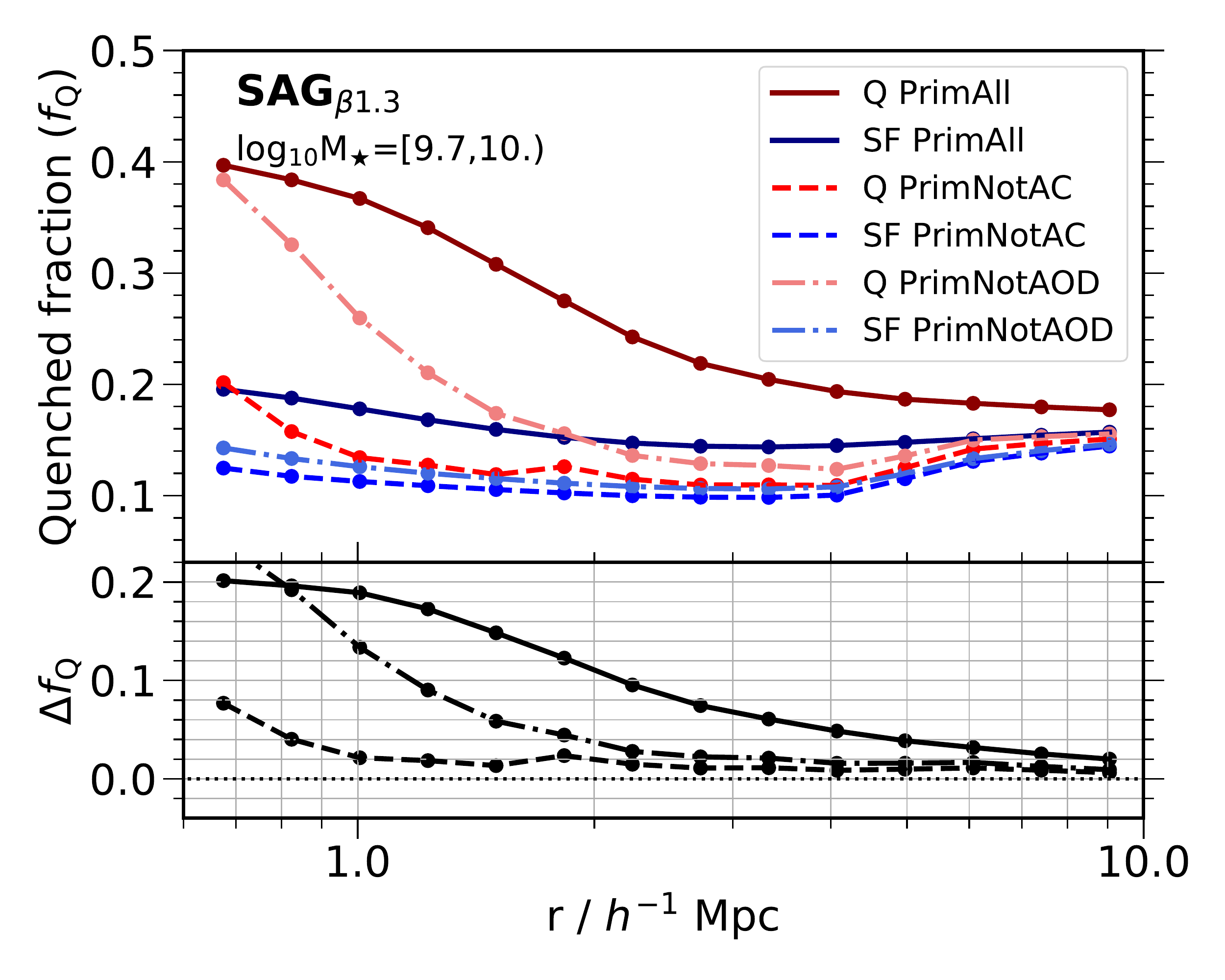}
\caption{Same as top panel in
Fig. \ref{fig_comp_ssfrBeta}, but 
including the \mbox{`PrimNotAOD'} sample based on the $\Sigma_5$ environmental estimator.
The dashed-dotted lines show the mean quenched fractions after removing from the primary sample the central galaxies in the vicinity of the densest systems in the 
\mdsagbeta~catalogue
(light coral and royal blue lines for quenched and star-forming primary galaxies, respectively, in the main panels).
}
\label{fig_ssfrBeta_AOD}
\end{figure}

For consistency with the exercise of Sec. \ref{sec_results} of using the most massive systems, we rank $\Sigma_5$
and select the systems with $\log_{10}(\Sigma_5 / h^2~\rm Mpc^{-2}) \geq$ 0.856 as proxies of the overdensities of interest.
With this cut, the number of selected overdensities 
equals
the number of haloes with masses  \mh\ $\geq$ 10$^{13}$  $\mhalo$ in \mdsagbeta.
Like in the sample `PrimNotAC',
we remove from the primary sample the central galaxies around overdensities (AOD) out to 5 $\mpc$;  
we refer to this sample as `PrimNotAOD'.
The results of the conformity for the 
`PrimNotAOD' sample
are shown in Fig. \ref{fig_ssfrBeta_AOD}
as dashed-dotted lines. 
They are different to the results obtained from the low-mass sample \mbox{`PrimNotAC'}
(also shown in this figure as dashed lines)
at distances $r \lesssim$ 2 $\mpc$ from the 
primary galaxies. 
The conformity signal is $\Delta f_{\rm Q}$  $\sim$ 0.14 and 0.04 at scales of 1 and 2 $\mpc$, respectively, when the low-mass central galaxies in the vicinity of overdensities are not considered in the primary sample, while $\Delta f_Q$ $\lesssim$ 0.02 at scales $\gtrsim$ 1 $\mpc$ for the
`PrimNotAC' sample.
Both cases are comparable for scales $r \gtrsim$ 2 $\mpc$, but the former always shows a slightly 
larger conformity signal than the latter. These results suggest that the environment around groups and clusters, defined by their halo mass, better characterizes conformity at a few Mpc than the environment around overdensities.

We also explore another definition of density environment. Instead of counting galaxies, we count subhaloes
in the {\sc IllustrisTNG300} simulation,
following the approach described in \citet{Artale+2018}, also implemented in \citet{Favole+2022}.
We use this environmental definition because it is an independent method for selecting overdensities, taking advantage 
of previously probed in this hydrodynamical cosmological simulation.
In this method, we count  
subhaloes with total mass above 
$10^{9.5}$ $\mhalo$ 
within a sphere of radius 3 $\mpc$, centred in dark matter haloes, divided 
by the volume of the sphere.
Unlike \citet{Artale+2018}, we also include the subhaloes that belong to the same host halo in which the sphere is centred.
The calculation is done by adopting periodic boundary conditions and is normalized by the number density of subhaloes in the box with the same mass cut.

\begin{figure}
\includegraphics[width=\columnwidth]{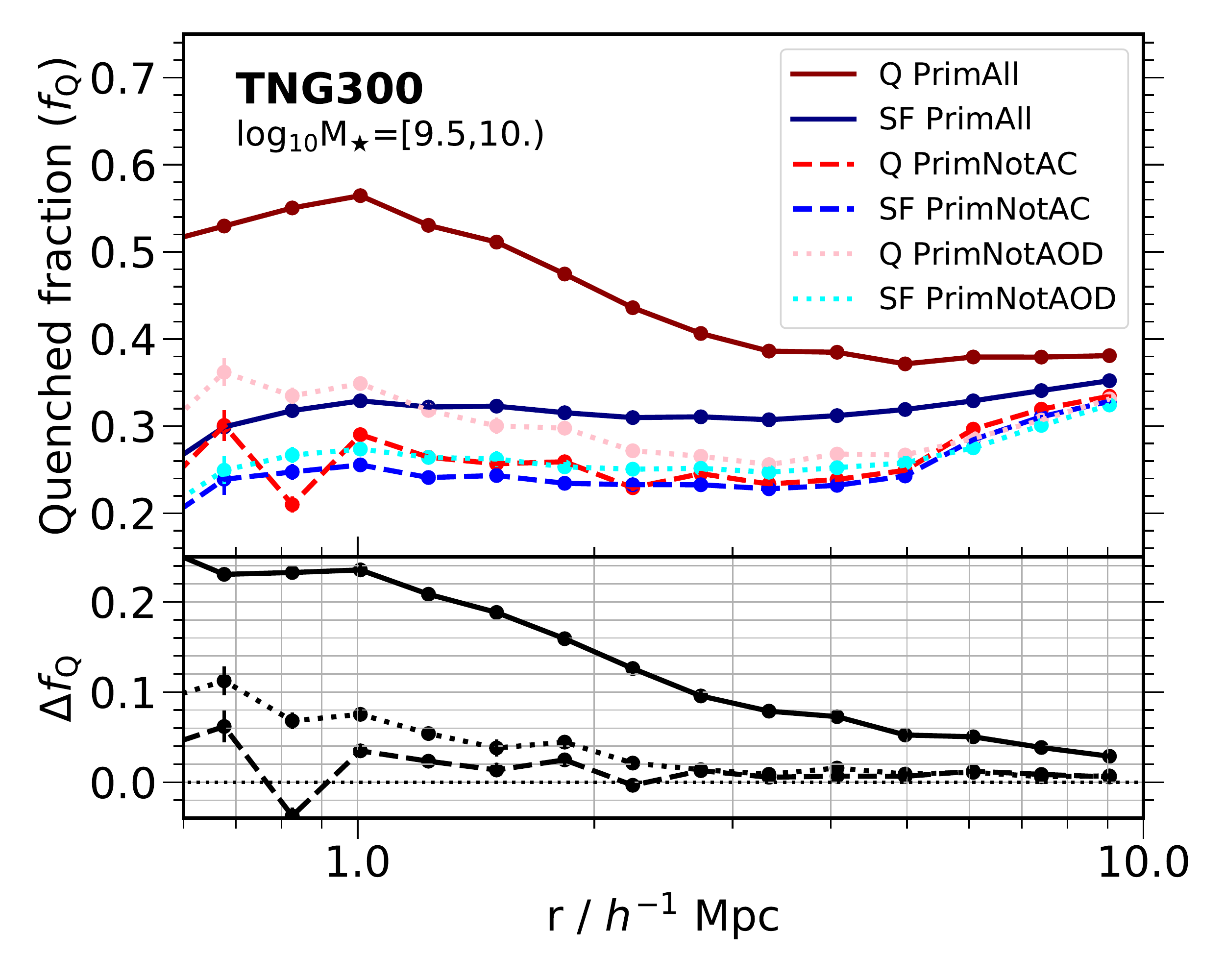}
\caption{Same as Fig.~\ref{fig_ssfrBeta_AOD}, but for primary galaxies in the stellar mass range 10$^{9.5}$ $\leq$ \ms/$\mstar$ $<$ 10$^{10}$ in the \tng{} simulation. In this case, results using an over-density estimator computed on scales of 3 $\mpc$ are included (dotted lines, see Sect. \ref{Sec_AOC} for details).
}
\label{fig_ssfrTNG300_AOD}
\end{figure}

We rank the systems using the environmental definition above to identify the highest densities as the overdensity regions of interest. 
For consistency, the number of 
the most overdense systems is the same as the number of haloes with \mh\ $\geq$ 10$^{13}$  $\mhalo$ in {\sc IllustrisTNG300}. 
We then remove from the primary sample the central galaxies around these overdensities out to 5 $\mpc$, which we refer again to as the 
sample `PrimNotAOD'. The mean quenched fractions of neighbours for 
this case
are shown in Fig.\ref{fig_ssfrTNG300_AOD} (dotted lines).
The conformity signal is $\Delta f_{\rm Q}$  $\sim$ 0.08 and 0.04 at scales of 1 and 2 $\mpc$, respectively, in the primary sample `PrimNotAOD'. These values are smaller than the case with the fiducial sample `PrimAll' (solid lines) with $\Delta f_{\rm Q}$  $\sim$ 0.24 and 0.14 at scales of 1 and 2 $\mpc$. The sample `PrimNotAC' (dashed lines) shows the smallest conformity signal at scales of $\sim$ 1 $\mpc$ with $\Delta f_{\rm Q}$ $\lesssim$ 0.04, but the signals between the cases `PrimNotAOD' and `PrimNotAC' are comparable from scales $r$ $\gtrsim$ 3 $\mpc$.
Again, these results suggest that it is better to characterize the systems that affect nearby low-mass central galaxies, responsible for the two-halo conformity, by the halo mass than the overdensities.

\subsection{Why is there a correlation between central galaxies in the vicinity of groups and clusters and neighbouring galaxies?}

In this paper, we have demonstrated 
an excess of correlation between quenched (red), low-mass central galaxies located in the vicinity of galaxy groups and galaxy
clusters and quenched (red) neighbouring galaxies out to distances of $r \sim$ 3 $\mpc$. 
The two-halo conformity signal
$\Delta f_{\rm Q}$ ($\Delta f_{\rm r}$) reduces to $\lesssim$ 0.02 when these low-mass central galaxies are not considered in the primary sample. This correlation mainly explains the two-halo conformity found in cosmological simulations and, probably, in observations. 
Why does this correlation exist?

Here, we make a simple test on checking the distance of quenched and star-forming central galaxies to the nearest massive halo with \mh\ $\geq$ 10$^{13}$ $\mhalo$ in \mdsag~catalogue.
For low-mass central galaxies with 
10$^{9.7}$ $\leq$ \ms/$\mstar$ $<$ 10$^{10}$,
the median distance is 1.3 $\mpc$ for quenched central galaxies, whereas it is 6.2 $\mpc$ for star-forming central galaxies. Therefore, quenched low-mass central galaxies are typically much closer to massive haloes than star-forming central galaxies of the same mass (by a factor of $\sim$ 5). 
For intermediate central galaxies with 10$^{10.3}$ $\leq$ \ms/$\mstar$ $<$ 10$^{10.5}$, the median distances to the nearest massive halo are similar between quenched and star-forming central galaxies (5.4 and 6.1 $\mpc$, respectively),
which suggests both populations tend to reside in similar large-scale environments relatively far from the influence of other massive haloes. For more massive central galaxies with 10$^{10.5}$ $\leq$ \ms/$\mstar$ $<$ 10$^{10.7}$, the median distances to the nearest massive halo are 5.5 $\mpc$ for quenched centrals and 6.2 $\mpc$ for star-forming centrals. These distances are very similar to those in the previous stellar mass bin. 
The combination of dominant internal processes in more massive galaxies (see Introduction) and the large separation from massive groups and clusters
can explain why the two-halo conformity signal is small ($\Delta f_{\rm Q}$ $\lesssim$ 0.02) for central galaxies at intermediate masses  and, also, for massive central galaxies.

The median distances of the star-forming central galaxies to the nearest massive halo are remarkably similar between the three stellar-mass ranges. On average, these galaxies continue their expected growth because they are far enough from the influence of massive systems.
In contrast, the median distance of low-mass quenched central galaxies to the nearest massive halo is notoriously smaller compared to those of more massive quenched central galaxies.
Our results agree with other works that show
the environmental effects are not locked within the virial radius only; they can extend further and affect nearby low-mass galaxies  \citep[e.g.][]{Wetzel+2012, Bahe+2013, Zinger+2018, Ayromlou+2021}. The two-halo conformity represents the environmental influence of relatively massive systems out to few Mpc scales.

It is not the scope of this paper to explore those environmental effects, which will be addressed elsewhere, but we can 
mention some of them studied in the literature. \cite{Bahe+2013} find that 
ram-pressure stripping exerted by the intracluster medium 
of simulated galaxy groups and clusters, which can be present beyond the virial radius of these systems \citep[e.g.][]{Zinger+2018}, can remove the hot gas content of low-mass galaxies ($\ms < 10^{10}$ $\msun$) out to $\sim$ 5 virial radii from the groups and clusters in the \textsc{GIMIC} cosmological hydrodynamic simulations. This ram-pressure stripping is much stronger for low-mass galaxies in filaments around clusters.
\cite{Zinger+2018} applied analytic models in simulated cluster systems and found that the ram-pressure stripping occurring in the outskirts of galaxy clusters may remove the satellite gas halo, but not the cold gas from the galaxy.
\cite{Ayromlou+2021} show that central galaxies can lose their hot gas via ram-pressure stripping in dense environments before infalling into a more massive halo using a novel version of the \textsc{L-Galaxies} SAM. This gas stripping is stronger for low-mass central galaxies near massive haloes.
\cite{Arthur+2019} find a causal link between the instantaneous ram pressure and the gas content of infalling haloes and subhaloes out to $\sim$ 1.5--2 virial radii from massive galaxy clusters in resimulated clusters of the \textsc{TheThreeHundred} project.
We note that, in the \sag~models, the ram-pressure stripping formalism is only implemented on satellite galaxies within the virial radius of host haloes.
This mechanism is not implemented on central galaxies in the outskirts of galaxy groups or clusters.

Any conformity signal obtained from the \sag~models in scales involving disconnected halo merger trees should be strictly driven by the different mass growth histories of the dark matter haloes because they depend on the environment. Indirect effects like backsplash galaxies give an alternative explanation. \citet{Bahe+2013} find that they 
have a significant contribution to the systematic depletion of cold gas out to $\sim$ 2--3 virial radii from simulated groups and clusters. However, these authors find that this indirect environmental effect is important for galaxies with $\ms > 10^{10}$ $\msun$.
On the other hand, 
it has been shown that relatively massive haloes could disrupt the expected growth of near smaller objects and, therefore, affect their properties
\citep[e.g.][]{WangH+2007, Dalal+2008, Hahn+2009, Salcedo+2018, MansfieldKravtsov2020},
which in turn may affect the properties of the central galaxies hosted by these disrupted haloes \citep[e.g.][]{LacernaPadilla2011}.
\citet{Behroozi+2014} find that the smooth accretion of mass of infalling haloes stops at median clustercentric distances of $\sim$ 1.8 virial radii of the final host halo at $z$ = 0.
Therefore, 
a feasible effect acting on the quenched, low-mass central galaxies is that the amount of gas accreted by their host dark matter haloes is strongly limited due to the stoppage in the growth of the host haloes produced by the nearby massive systems.
Without gas replenishment, star formation in these low-mass central galaxies is halted.

The correlation between central galaxies in the vicinity of groups and clusters and neighbouring galaxies is because the environment around these massive systems has probably affected the expected growth of nearby haloes and, consequently, of the central galaxies hosted by them. The nearby truncated, low-mass central galaxies are typically quenched and exhibit red colours, similar to satellite galaxies inside the virial radius of massive systems. 
%

\section{Conclusions}
\label{sec_conclusion}

We have studied the environment 
around groups and clusters of galaxies with halo mass \mh\ $\geq$ 10$^{13}$ $\mhalo$ using two 
galaxy catalogues generated from 
different versions of 
the semi-analytic model {\sc sag} applied to the {\sc mdpl2} cosmological simulation (\mdsag~and \mdsagbeta~catalogues), 
and the \tng{} cosmological hydrodynamical simulation.
Low-mass central galaxies in the vicinity
of these massive systems out to 5 $\mpc$ are preferentially quenched compared to other central galaxies at fixed stellar mass or fixed host halo mass at $z \sim 0$ in all these cosmological simulations.
We find consistently in all the 
galaxy catalogues that these low-mass central galaxies, especially those in the vicinity of galaxy groups,
mostly produces the two-halo galactic conformity measured at
large separations of several Mpc between the low-mass central galaxy and neighbouring galaxies in adjacent haloes.

In summary, we measure the mean quenched fractions of neighbouring galaxies as functions of the real 
space distance from quenched primary galaxies and star-forming primary galaxies. The primary samples correspond to central galaxies in the simulations. 
The galactic conformity signal for a given distance, $\Delta f_{\rm Q}$($r$), is measured as the difference between these mean quenched fractions at fixed mass of the primary galaxies. For the fiducial sample `PrimAll', the conformity is important for low-mass primary galaxies (\ms $\leq$ 10$^{10}$ $\mstar$), with $\Delta f_{\rm Q}$ $\sim0.15$, $\sim 0.19$, and $\sim0.24$ at $r$ $\sim$ 1 $\mpc$ in the 
\mdsag, \mdsagbeta, 
and the \tng{} simulation, respectively. The conformity signal declines to $\Delta f_{\rm Q}$ values lower than 0.02 at distances larger than 7 $\mpc$ from the low-mass primary galaxies in these simulations (Figs. \ref{fig_comp_ssfr}, \ref{fig_comp_ssfrBeta}, and \ref{fig_fQ_IllustrisTNG}). 
The conformity signal is significantly reduced when the central galaxies in the vicinity of groups and clusters are removed from the primary
sample (`PrimNotAC')
with $\Delta f_{\rm Q}$ $\lesssim$ 0.02 for scales $r \gtrsim$ 1 $\mpc$ in both SAMs (Figs. \ref{fig_comp_ssfr} and  \ref{fig_comp_ssfrBeta}), and $\Delta f_{\rm Q}$ $\lesssim$ 0.02 at $r \gtrsim$ 1.5 $\mpc$ in \tng{} (Fig. \ref{fig_fQ_IllustrisTNG}). 
The low-mass central galaxies in the vicinity of low-mass groups are responsible for the majority of the conformity signal (Fig. \ref{fig_fQnotAC_IllustrisTNG}).

The trends with the conformity signal remain for primary galaxies at fixed halo mass. For primary galaxies in low-mass host haloes of 10$^{11.4}$ $\leq$ \mh/$\mhalo$ $<$ 10$^{11.6}$ in the \mdsag~catalogue,
$\Delta f_{\rm Q}$ $\sim$ 0.16 at $r \sim$ 1 $\mpc$ 
in the sample `PrimAll',
and it reduces to $\Delta f_{\rm Q}$ $\sim$ 0.02 at scales larger than 7 $\mpc$. However, $\Delta f_{\rm Q}$ is $\lesssim$ 0.02 at distances $r \gtrsim$ 1 $\mpc$ from the primary galaxies in the 
sample `PrimNotAC'
(Fig. \ref{fig_comp_ssfr_fixedMhalo}).

We tested that the results in the sample `PrimNotAC' are not an artefact of removing galaxies in the primary sample.  Removing central galaxies randomly from the primary sample does not modify the quenched fractions of neighbours at distances of
a few Mpc from the primary galaxies in ‘PrimAll’ 
(Fig. \ref{fig_testA_TNG300}).
The conformity signal at few Mpc scales depends on the vicinity radius around relatively massive haloes in which we remove central galaxies, though (Fig. \ref{fig_diff_Mh_R}), but we tested that this influence on low-mass central galaxies does not reach scales much more extensive than 5 $\mpc$.

We also explored the effects in central galaxies around overdensities (`PrimNotAOD') instead of the case around massive haloes. By using two definitions for selecting the densest systems, the case `PrimNotAOD' is comparable to the case `PrimNotAC' for
scales $r \gtrsim$ 2--3 $\mpc$, but the environment around groups and clusters defined by their halo mass better characterizes the two-halo conformity than the environment around overdensities at $r \gtrsim$ 1 $\mpc$, in general (Figs. \ref{fig_ssfrBeta_AOD} and \ref{fig_ssfrTNG300_AOD}).

The quenched, low-mass central galaxies are much closer to massive haloes than star-forming central galaxies of the same mass (by a factor of $\sim$ 5). Future works will be needed to determine if the host haloes of the quenched, low-mass central galaxies have been disrupted by the overwhelming presence of nearby massive haloes, which in turn may affect the expected star formation of these low-mass central galaxies.

\section*{Acknowledgements}
The authors acknowledge the kind support of the computing team at IATE.
The authors thank the anonymous referee for the revision that helped improve the presentation of this paper.
FR, ALO and ANR thanks the
support by {\it Agencia Nacional de Promoción Científica y Tecnológica, 
Consejo Nacional de Investigaciones
Científicas y Técnicas} (CONICET, Argentina), and 
{\it Secretaría de Ciencia y Tecnología de la Universidad Nacional de Córdoba}
(SeCyT-UNC, Argentina).
ADMD thanks Fondecyt for financial support through the Fondecyt Regular 2021 grant 1210612.
SAC acknowledges funding from {\it Consejo Nacional de Investigaciones Cient\'{\i}ficas y T\'ecnicas} (CONICET, PIP-0387), {\it 
Agencia Nacional de Promoci\'on de la Investigaci\'on, el Desarrollo Tecnol\'ogico y la Innovaci\'on} (Agencia I+D+i, PICT-2018-3743), and {\it Universidad Nacional de La Plata} (G11-150), Argentina.
MCA acknowledges financial support from the Austrian National Science Foundation through FWF stand-alone grant P31154-N27.
CVM acknowledges support from ANID/FONDECYT through grant 3200918, and he also acknowledges support from the Max Planck Society through a Partner Group grant.
The authors gratefully acknowledge the Gauss Centre for Supercomputing e.V. (www.gauss-centre.eu) and the Partnership for Advanced Supercomputing in Europe (PRACE, www.prace-ri.eu) for funding the MultiDark simulation project by providing computing time on the GCS Supercomputer SuperMUC at Leibniz Supercomputing Centre (LRZ, www.lrz.de).

\section*{Data Availability}
The data underlying this article is available as follows. 
The \mdsag{} galaxy catalogue is publicly available at the CosmoSim database
http://www.cosmosim.org/.
The \mdsagbeta{} galaxy catalogue will be shared on reasonable request to the corresponding author.
The \tng{} simulation is publicly available at the \textsc{TNG} website https://www.tng-project.org/.

\bibliography{references}

\appendix

\section{Conformity with colour}
\label{sec_colour}

\begin{figure}
\includegraphics[width=\columnwidth]{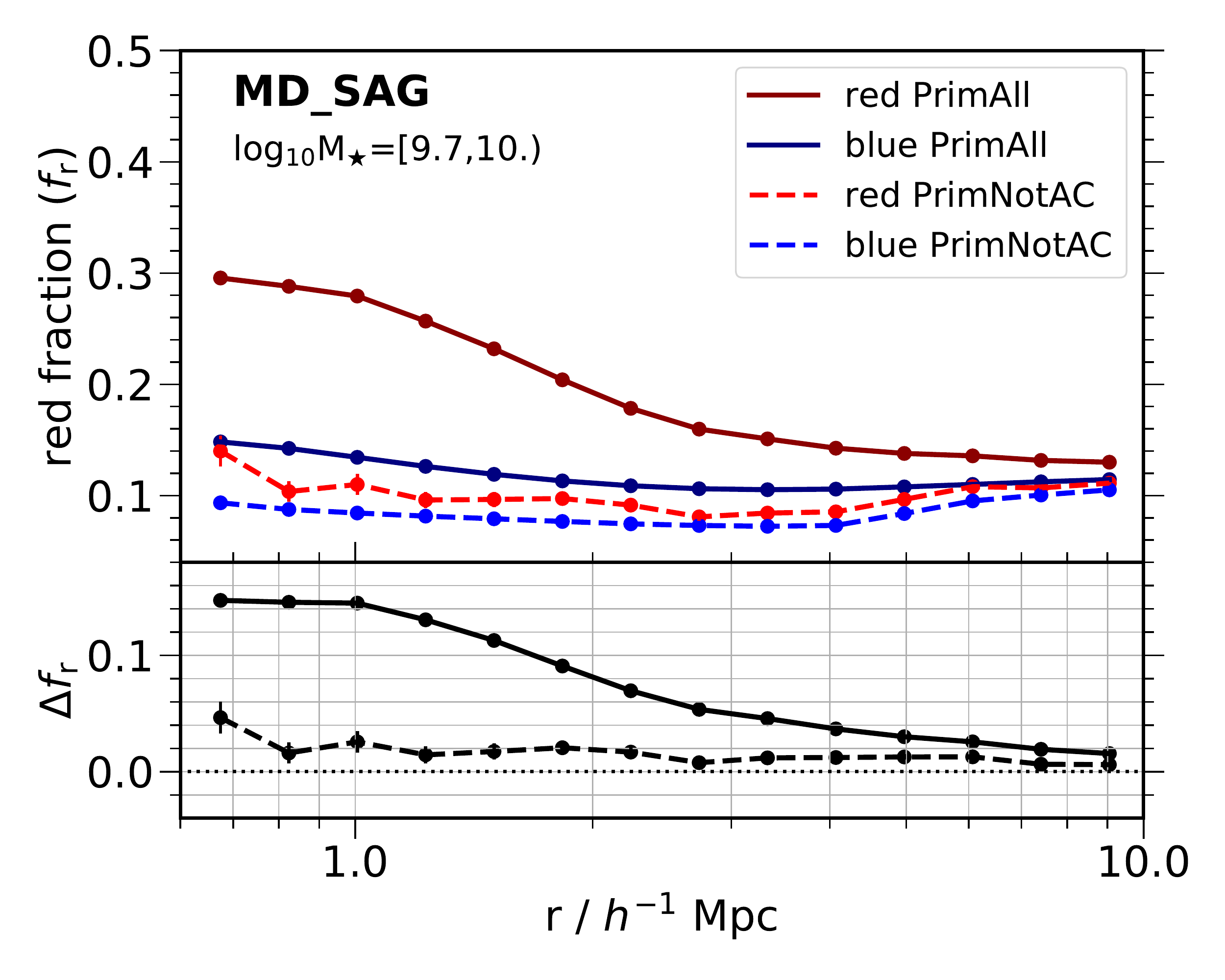}
\includegraphics[width=\columnwidth]{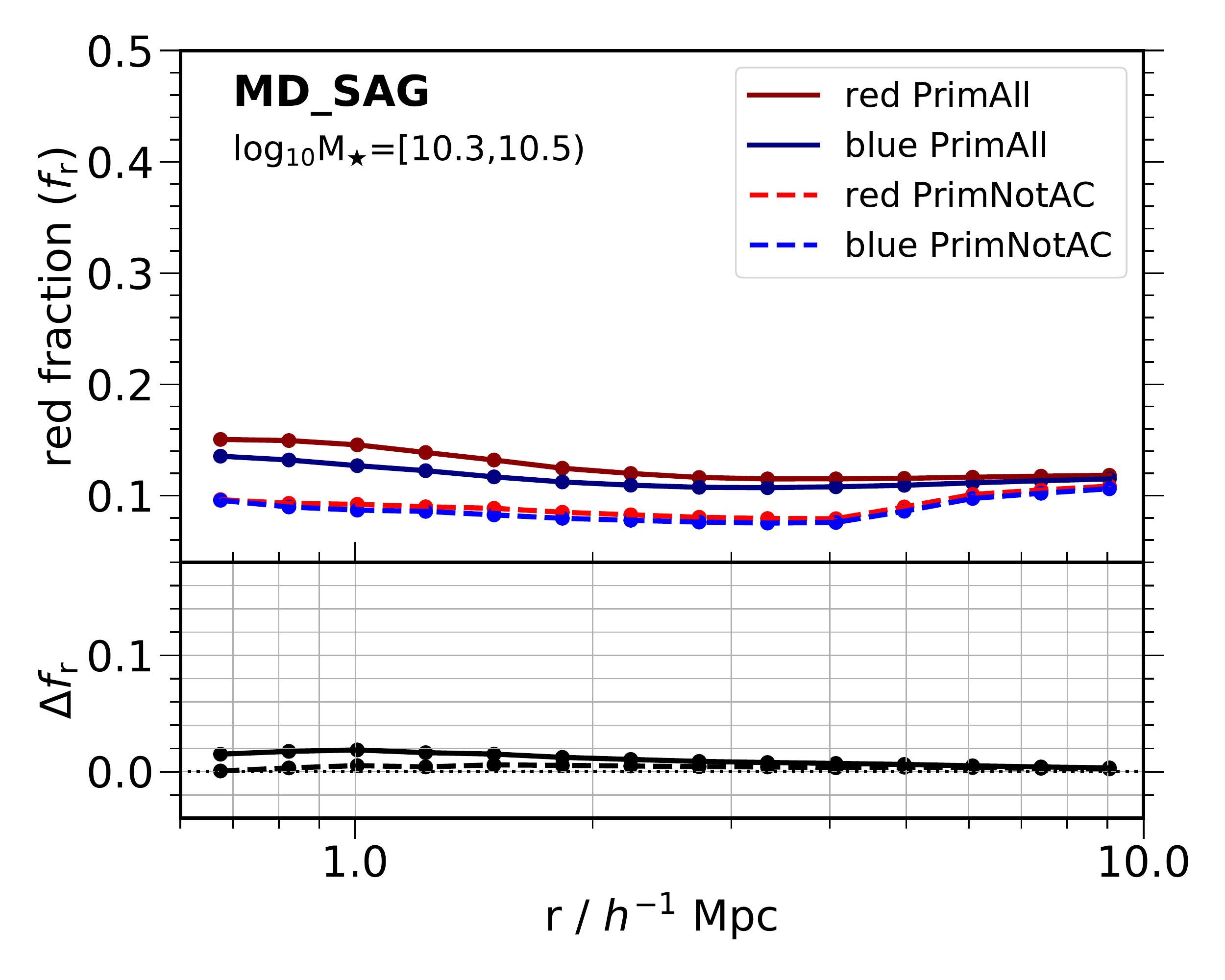}
\caption{Similar as Fig. \ref{fig_comp_ssfr} but using $g - r$ colour instead of sSFR.
}
\label{fig_comp_gr}
\end{figure}

We explore the two-halo conformity trends using the $g - r$ colour.
Fig. \ref{fig_comp_gr} shows the mean red fractions of neighbouring galaxies as functions of the distance from the primary galaxies in two stellar-mass bins in the 
\mdsag~galaxy catalogue.
The solid lines correspond to the case `PrimAll'. 
The correlation in the red colour between neighbours and primary galaxies at distances of a few Mpc is much stronger for low-mass primaries (10$^{9.7}$ $\leq$ \ms/$\mstar$ $<$ 10$^{10}$) 
compared with primaries at intermediate masses (10$^{10.3}$ $\leq$ \ms/$\mstar$ $<$ 10$^{10.5}$). 
The differences in the mean red fractions of neighbours around red and blue primary galaxies, $\Delta f_{\rm r}$, are shown with black solid lines in the sub-panels. For the low-mass primaries, $\Delta f_{\rm r}$ is as big as $\sim$ 0.14 at $\sim$ 1 $\mpc$, it decreases to $\sim$ 0.05 at $\sim$ 3 $\mpc$, and it is smaller than 0.02 at distances $\gtrsim$ 7 $\mpc$. For primary galaxies of intermediate masses, $\Delta f_{\rm r}$ is typical $\lesssim$ 0.02 at all the scales. 
These results are both qualitatively and quantitatively very similar to the difference in the mean quenched fractions (black solid lines in Fig. \ref{fig_comp_ssfr}).
Therefore, there is a clear correlation between neighbouring galaxies and low-mass primary galaxies in both the colour and sSFR  out to distances of a few Mpc.

The case `PrimNotAC', in which we remove from the primary sample the central galaxies in the vicinity of groups and clusters, is shown as dotted lines in Fig. \ref{fig_comp_gr}. 
The difference in the mean red fractions of the case `PrimNotAC' is typical $\Delta f_{\rm r}$ $\lesssim$ 0.03 at distances $\gtrsim$ 1 $\mpc$ from low-mass primary galaxies, and it is smaller for the intermediate-mass primary galaxies. For the low-mass primary galaxies, $\Delta f_{\rm r}$ decreases a factor of 3--4 at 2--3 $\mpc$ in `PrimNotAC' compared with \mbox{`PrimAll'}.
The overall behaviour of the mean red fractions in the case `PrimNotAC' is qualitatively very similar to the same case of the mean quenched fractions (dotted lines in Fig. \ref{fig_comp_ssfr}). Therefore, 
the observed correlation in both colour and sSFR between low-mass central galaxies and neighbour galaxies at distances of a few Mpc is mainly driven by the central galaxies located in the outskirts of groups and clusters of galaxies.

Taking into account the remarkable similarity between the results obtained from the 
\mdsag{} and \mdsagbeta{} models, and the \tng{} simulation when considering the sSFR, it is reasonable to expect that the same trend is maintained in the \mdsagbeta{} and \tng{} galaxy catalogues
when considering colours. Indeed, this is the case, but we do not show such analysis here.

\label{lastpage}
\end{document}